\documentclass[sigconf, nonacm]{acmart}
\usepackage{xspace}
\usepackage{enumitem}
\usepackage{multirow}
\usepackage{colortbl}
\usepackage[ruled,vlined]{algorithm2e}
\usepackage{pvldb}

\renewcommand\vldbdoi{XX.XX/XXX.XX}
\renewcommand\vldbpages{XXX-XXX}

\renewcommand\vldbavailabilityurl{https://anonymous.4open.science/r/SemanticFilter} 

\begin{document}

\newcommand{\bao}[1]{{ \textcolor{blue}{\textsf{Bao says: #1}}}}
\newcommand{\zepeng}[1]{{ \textcolor{orange}{\textsf{Zepeng says: #1}}}}
\newcommand{\hai}[1]{{ \textcolor{violet}{\textsf{Hai says: #1}}}}
\newcommand{\tocheck}[1]{\textcolor{red}{#1}}
\newcommand{\toread}[1]{\textcolor{blue}{#1}}
\newcommand{\ul}[1]{\underline{#1}}
\newcommand{\todo}[1]{\textcolor{green}{TODO: #1}}
\newcommand{\myparagraph}[1]{\vspace{0.3\baselineskip}\noindent{\textbf{#1.}}~}
\newcommand{\norm}[1]{\left\lVert #1\right\rVert_2}

\newcommand{\aurora}{\texttt{Falcon}\xspace}
\newcommand{\lotus}{\texttt{LOTUS}\xspace}
\newcommand{\bargain}{\texttt{BARGAIN}\xspace}
\newcommand{\scaledoc}{\texttt{ScaleDoc}\xspace}
\newcommand{\csv}{\texttt{CSV}\xspace}

\newcommand{\bank}{\texttt{Banking77}\xspace}
\newcommand{\emotion}{\texttt{GoEmotions}\xspace}
\newcommand{\patent}{\texttt{BigPatent}\xspace}
\newcommand{\law}{\texttt{LeCaRDv2}\xspace}
\newcommand{\med}{\texttt{PubMed}\xspace}
\newcommand{\mimic}{\texttt{MIMIC-III}\xspace}

\newcommand{\auroradoc}{\texttt{Falcon}-\texttt{Doc}\xspace}

\title{\texttt{Falcon}: Fast, Lightweight, and Robust Cascade for Oracle-Efficient Semantic Filtering}

\author{Zepeng Liu$^\dagger$}
\affiliation{%
  \institution{The University of Queensland}
  \state{Brisbane}
  \country{Australia}
}
\email{notnnorth@gmail.com}

\author{Hai Lan$^\dagger$}
\affiliation{%
  \institution{The University of Queensland}
  \state{Brisbane}
  \country{Australia}
}
\email{h.lan@uq.edu.au}

\author{Yiming Shi}
\affiliation{%
  \institution{The University of Queensland}
  \state{Brisbane}
  \country{Australia}
}
\email{yimingshi666@gmail.com}

\author{Sheng Wang}
\affiliation{%
  \institution{Wuhan university}
  \city{Wuhan}
  \country{China}
}
\email{swangcs@whu.edu.cn}

\author{Zhifeng Bao}
\affiliation{%
  \institution{The University of Queensland}
  \state{Brisbane}
  \country{Australia}
}
\email{zhifeng.bao@uq.edu.au}


\begin{abstract}
Natural-language predicates allow users to filter documents using complex semantic criteria rather than rigid rules in structured queries, making them increasingly vital for modern data analytics. 
However, evaluating these predicates with a large language model on every document is prohibitively expensive. A proxy cascade reduces this cost by resolving confident documents with a cheap proxy and routing only uncertain ones to the LLM oracle. Yet, existing proxies score documents by predicate-agnostic content, waste labels under extreme selectivity, and select conservative thresholds that result in unnecessary oracle calls.
We present \aurora, a semantic filter that minimizes oracle calls while certifying a target quality with high probability. \aurora transforms generic embeddings into predicate-aware semantic vectors using LLM-derived priors and a small oracle-labeled sample. This sample is drawn via an information-guided sampling that remains effective even when positives are rare. \aurora then scores each document with a lightweight multi-view proxy and selects the most aggressive cascade thresholds, whose quality is certified through an additive F1 budget. Experiments on six datasets show that \aurora reduces end-to-end latency and token cost by up to $7.3\times$ over existing methods while meeting the target F1.
\end{abstract}

\maketitle

\vldbtopmatter

\vspace{-0.25em}
\section{Introduction}
Large language models (LLMs)~\cite{GPT} 
enable semantic operators over documents, such as joins~\cite{zeighami2025featurized,trummer2026optimal,Patel25lotus}, filters~\cite{hou26csv,Patel25lotus,Zeighami25bargain,zhang2026scaledoc}, and sorts~\cite{zhao2025access,Patel25lotus}. This work focuses on \emph{semantic filtering}: retrieving documents that satisfy a natural-language predicate.
For example, a patent analyst may retrieve documents that describe ``a new chemical composition,'' while an enterprise may identify customer reviews that ``express dissatisfaction with service quality''. The naive approach evaluates each document with a strong LLM. While accurate, it is prohibitively expensive for large collections. 

To reduce cost, prior studies~\cite{Patel25lotus,Zeighami25bargain,hou26csv,zhang2026scaledoc} adopt \emph{selective oracle invocation}: a cheaper proxy resolves high-confidence documents, routing only the remainder to a strong LLM oracle. 

\vspace{-0.8em}
\subsection{Identified Research Gaps}
Three lines of work instantiate the above paradigm: (1) LLM-based proxy cascades, e.g., \lotus \cite{Patel25lotus} and \bargain \cite{Zeighami25bargain}, (2) learned proxy models with threshold selection, e.g., \scaledoc \cite{zhang2026scaledoc}, and (3) clustering-based proxies, e.g., \csv \cite{hou26csv}. 
Yet, three research gaps remain: 

\noindent\textbf{Gap 1: Proxy signals are either costly or weakly discriminative}.
LLM-based proxies~\cite{Patel25lotus,Zeighami25bargain} preserve predicate semantics 
but incur one expensive model call per document, limiting their scalability to large collections.
Learned proxies~\cite{zhang2026scaledoc} avoid these calls by scoring pretrained document embeddings, but these embeddings also contain predicate-irrelevant information, introducing noise that complicates learning and weakens discrimination between positive and negative documents.

\noindent\textbf{Gap 2: Uniform label acquisition is ineffective under extreme selectivity}.
Learned 
\cite{zhang2026scaledoc} and clustering-based proxies 
\cite{hou26csv} rely on uniformly sampled oracle labels. 
When one class is rare (i.e., extreme selectivity), a limited uniform sample may contain too few or none of its documents. Hence, the learned proxy lacks rare-class examples needed to estimate a stable decision boundary. And cluster-quality estimates are supported by insufficient class evidence and have high variance. 

\noindent\textbf{Gap 3: Quality-constrained routing is either unsafe or overly conservative}. A proxy-first, oracle-fallback cascade 
requires estimating the collection-level quality from only a small labeled sample. Loose estimates force conservative thresholds, routing too many documents to the oracle. Aggressive estimates risk violating the target quality. This gap amplifies when non-uniform sampling is used to address Gap 2.

\vspace{-0.8em}
\subsection{Our Solution and Contributions}
To mitigate these gaps, we propose \aurora, a \underline{\textbf{fa}}st, \underline{\textbf{l}}ightweight, and robust \underline{\textbf{c}}ascade for \underline{\textbf{o}}racle-efficie\underline{\textbf{n}}t semantic filtering. \aurora's core idea is: 
\emph{progressively transform a limited oracle-label budget into predicate-specific evidence, using this evidence to construct a discriminative proxy and select cascade thresholds under a quality constraint}. 
This leads to three tightly connected components.

\noindent$\bullet$~\textbf{Predicate-Aware Semantic Vectors
(Section~\ref{sec:pred-vec-gen-main}).}
As highlighted in \textbf{Gap 1}, a low-cost proxy must capture predicate-specific semantics without per-document LLM calls. However, pretrained embeddings contain predicate-irrelevant variation that can obscure discriminative signals. \aurora uses a single LLM call to generate positive and negative semantic descriptions as candidate discriminative directions. Guided by these directions, it samples both score tails to improve rare-class coverage, retains directions supported by oracle labels, and suppresses irrelevant corpus-level variation. The resulting compact semantic vectors capture predicate-specific information without per-document LLM inference.

\noindent$\bullet$~\textbf{Lightweight Proxy with Progressive Refinement (Section~\ref{sec:light-proxy-model}).}
Predicate-aware representations do not ensure reliable routing: a cascade requires scores that separate positive and negative documents, particularly at decision extremes. As highlighted in \textbf{Gap 2}, learning such scores from limited labels is challenging under extreme selectivity, where rare-class evidence is scarce. Rather than training a high-capacity classifier \cite{zhang2026scaledoc}, \aurora combines directional, neighborhood, and semantic-prior signals in a lightweight multi-view proxy, with residual correction for systematic scoring errors. Initial tail-guided sampling improves rare-class coverage, while subsequent uncertainty-guided refinement targets overlapping score regions where the proxy remains unreliable. Newly acquired labels update both the semantic vectors and proxy, progressively improving discrimination under limited supervision.


\noindent$\bullet$~\textbf{Quality-Certified Threshold Selection (Section~\ref{sec:th-select-wg}).}
As highlighted in \textbf{Gap 3}, even a discriminative proxy cannot guarantee that aggressive routing satisfies the target F1. Estimating this collection-level metric from limited, non-uniformly sampled labels introduces uncertainty, leading to conservative or unsafe thresholds. \aurora addresses this by reformulating the F1 constraint as an equivalent additive document-level budget. It uses a score-stratified pilot to identify promising threshold pairs, then allocates additional audit samples based on candidate routing behavior and estimated uncertainty. This enables simultaneous lower confidence bounds on quality budgets, allowing \aurora to select the highest-reduction candidate whose bound meets the target. Under the stated regularity conditions, the quality guarantee holds asymptotically.


\noindent$\bullet$~\textbf{Extensive Experiments (Section~\ref{sec:experiment})}. 
We evaluate \aurora on six real-world datasets using \(120\) semantic-filtering predicates with different selectivities and target F1 constraints. Compared with five representative baselines, \aurora reduces oracle calls by up to 82\% and end-to-end latency and token consumption by up to $7.3\times$, while closely meeting the target F1. Ablation studies further demonstrate the effectiveness of each major component.

\section{Problem Formulation and Overview}

\subsection{Problem Definition}

We consider a document collection \(\{d_i\}_{i=1}^{n}\) and a binary predicate \(p\). Each document $d_i$ has an  oracle label $y_i$ given by an oracle labeler $\mathcal{L}_o$, i.e., $y_i=\mathcal{L}_o(d_i,p)\in\{0,1\}$. \(y_i=1\) indicates that \(d_i\) satisfies \(p\), and \(y_i=0\) otherwise. We refer to documents with $y_i=1$ and $y_i=0$ as positive and negative documents, respectively.
Following prior studies~\cite{zhang2026scaledoc,Patel25lotus,Zeighami25bargain}, we instantiate the oracle, $\mathcal{L}_o$, with a strong LLM. Under this setting, a semantic filter produces a predicted label $\hat{y}_i\in\{0,1\}$ for every document. It may invoke the oracle on a subset \(D'\subseteq D\) and predict the labels of the remainder without oracle evaluation. We formally define the studied problem as follows.

\begin{definition}[Quality-constrained Semantic Filtering Problem \cite{zhang2026scaledoc}]~\label{def:sem_filter}
Given a document collection $\{d_i\}_{i=1}^{n}$, a binary predicate $p$, a quality metric $\mathcal{Q}$, a target quality level $\alpha$, and a failure probability $\delta$, quality-constrained semantic filtering seeks a predicted labeling $\{\hat{y}_i\}_{i=1}^{n}$ over $D$ for $p$ and an oracle-routed set \(D'\) that solve:
\begin{equation}
\min \ \frac{|D'|}{|D|} \quad \text{s.t.} \quad \mathbb{P}[\mathcal{Q}(y, \hat{y}) \geq \alpha] \geq 1 - \delta.
\end{equation}
\end{definition}

Routing every document to the oracle trivially yields $\hat{y}_i=y_i$, i.e., \(D'=D\). The challenge is to minimize oracle calls while preserving the required quality guarantee.

\begin{table}[t] \small
\centering
\caption{Summary of Notations.}
\vspace{-1em}

\label{tab:notation}
\setlength{\tabcolsep}{2.4pt}
\scalebox{1.03}{\begin{tabular}{cc}  
\toprule
\textbf{Notations} & \textbf{Description} \\
\midrule
\(D=\{1,2,\dots,n\}\) & The identifier collection of \(n\) documents. \\
\(p\) & Target binary predicate over documents. \\
\(y_i, \hat{y_i} \in \{0,1\}\) & Oracle and predicted labels of the \(i\)-th document \(d_i\). \\
\(\mathbf{x}_i, \mathbf{z}_i\) & Original embedding and semantic vectors of \(d_i\). \\
\(\mathbf{a}^+,\mathbf{a}^-\) & Positive and negative anchor descriptions. \\
\(P^+,P^-\) & Positive and negative facet collections. \\
\(S\) & Oracle-labeled sampled document-id set. \\
\(\mathbf{C}_p,\mathbf{N}_p\) & Predicate task basis and nuisance basis. \\
\(\ell_i\) & Proxy score assigned to document \(d_i\). \\
\(l,r\) & Lower and upper decision thresholds. \\
\(\alpha,\delta\) & Target quality level and failure probability. \\
\(\norm{\cdot}\) & The Euclidean norm. \\
\(\phi(\textbf{x})=\textbf{x}/\norm{\textbf{x}}\) & Vector normalization for vector \(\textbf{x}\ne\mathbf{0}\). \\
\(\mathcal{R}(\cdot)\) & The min-max rank map. \\
\bottomrule
\end{tabular}}
\vspace{-1em}
\end{table}

Table~\ref{tab:notation} summarizes the notations used throughout the paper.

\subsection{Proxy Cascade in \aurora}\label{sec:proxy_casc}
Like existing cascade-based approaches~\cite{Patel25lotus,Zeighami25bargain,zhang2026scaledoc}, \aurora instantiates quality-constrained semantic filtering as a \emph{proxy-first, oracle-fallback cascade}, which we refer to as a \emph{proxy cascade}. The proxy assigns each document $d_i$ a proxy score $\ell_i \in [0, 1]$, where a larger score indicates stronger evidence that $d_i$ satisfies the predicate $p$.

Given two thresholds $0 \leq l < r \leq 1$, \aurora resolves documents with $\ell_i \le l$ as negative and those with $\ell_i \ge r$ as positive. The remaining documents form the oracle-routed set $D'(l, r) = \{d_i : l < \ell_i < r\}$ and their labels are obtained from the oracle. Let $\hat{y}(l,r)$ denote the complete labeling produced by this cascade. The fraction $|D'(l, r)| / |D|$ is the oracle-call fraction while \(1-|D'(l,r)|/|D|\) is the oracle-call reduction.

Under this cascade, the problem in Definition~\ref{def:sem_filter} reduces to selecting the threshold pair \((l,r)\) that minimizes the oracle-call fraction while satisfying the quality guarantee:
\begin{equation}\label{eq:our_goal}
\min_{l, r} \ \frac{|D'(l, r)|}{|D|} \quad \text{s.t.} \quad \mathbb{P}[\mathcal{Q}(y, \hat{y}(l,r)) \geq \alpha] \geq 1 - \delta.
\end{equation}

We use the F1 as the quality metric $\mathcal{Q}$ by default. 
\aurora also supports precision or recall by replacing the quality constraint.



\section{Predicate-Aware Semantic Vector Generation}\label{sec:pred-vec-gen-main}

\begin{figure}[t]
  \centering
  \includegraphics[width=0.48\textwidth]{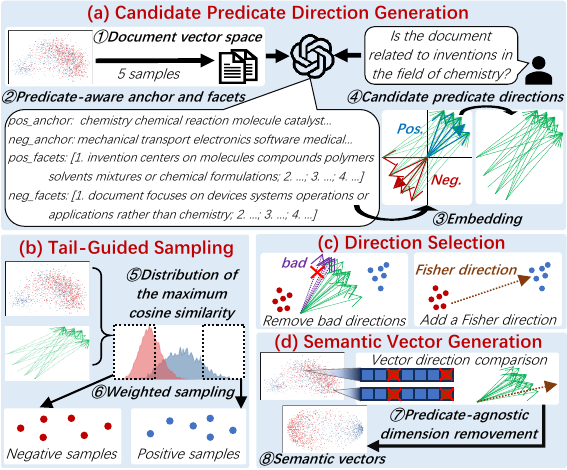}
  \vspace{-2.00em}
  \caption{Predicate-aware semantic vector generation.}
  \vspace{-1.00em}
  \label{fig:section3}
\end{figure}

To reduce oracle calls, the proxy scores must clearly separate positive and negative documents. Pretrained document embeddings are not ideal because they primarily capture general content similarity rather than predicate satisfaction. 
Our solution is to transform each pretrained embedding into a compact, predicate-aware semantic vector. Constructing such vectors under a small oracle-label budget faces three challenges: 1) the aspects that distinguish positive from negative documents are unknown in advance; 2) uniform sampling may yield too few examples from the rare class; 3) learning a transformation directly from a small labeled sample may overfit.

\aurora addresses these challenges in four steps: 1) It first uses a single LLM call and five oracle-labeled documents to generate candidate predicate directions (Figure~\ref{fig:section3} (a)) in Section~\ref{sec:pred-direction}. 2) It then applies tail-guided sampling to increase the chance of acquiring both positive and negative documents (Figure~\ref{fig:section3} (b)) in Section~\ref{sec:tail-samp}. 3) \aurora retains the candidate directions supported by the acquired labels and adds a label-derived direction (Figure~\ref{fig:section3} (c)) in Section~\ref{sec:di-select}. 4) It removes corpus-level variation unrelated to the predicate and projects each document into a predicate-aware semantic vector (Figure~\ref{fig:section3} (d)) in Section~\ref{sec:pred-vec-gen}.


\subsection{Candidate Predicate Direction}\label{sec:pred-direction}

Let $\mathbf{x}_i=\mathcal{E}(d_i)\in \mathbb{R}^m$ denote the pretrained embedding of document $d_i$, where $\mathcal{E}$ is the embedding function. 
A predicate direction is a unit vector that points from a negative description toward a positive one. A document's projection onto this direction indicates its alignment with the positive class. Since the correct directions are unknown, \aurora first generates a set of candidates.

Prior studies suggest that LLMs can generate useful semantic cues~\cite{zhou2023flame,kuhn2023semantic,arora2023evaporate}, while abstract concepts can admit linear representations~\cite{sofroniew2026emotion}. Inspired by these findings, we use differences between positive and negative description embeddings as candidate predicate directions, pending validation against oracle labels. For a predicate $p$, \aurora randomly selects five documents, obtains their oracle labels, and prompts an LLM once with $p$ and the labeled examples. These examples provide the domain and wording of the collection and need not cover both classes. The LLM generates two types of descriptions, namely the positive and negative anchors that summarize the two classes, together with facets describing finer class-specific criteria. Figure~\ref{fig:section3}(a) shows an example.

When a user asks, ``\emph{Is the document related to inventions in the field of chemistry?}'', the LLM refers to the expression style of the sampled documents and generates keywords that it considers indicative of predicate satisfaction or violation, such as \emph{chemical reaction} as positive anchors, and \emph{mechanical} and \emph{transport} as negative anchors. In addition, the LLM generates several facets as guidance for positives and negatives, as shown in step~\textcircled{2} of Figure~\ref{fig:section3}(a).

\aurora embeds all generated descriptions with $\mathcal{E}$. Let $\mathbf{a}^+$ and $\mathbf{a}^-$ denote the positive and negative anchor embeddings. Let $P^{+}$ and $P^{-}$ denote the sets of positive and negative facet embeddings. The anchor pair defines one candidate direction, $\mathbf{v}_a=\phi(\mathbf{a}^+-\mathbf{a}^-)$. Each positive–negative facet pair defines another, $\mathbf{v}_u=
  \phi(\mathbf{u}^+-\mathbf{u}^-), (\mathbf{u}^+,\mathbf{u}^-)\in P^{+}\times P^{-}$,
where $\phi(\textbf{x})=\textbf{x}/\norm{\textbf{x}}$ normalizes a vector to unit length. Together, the anchor and facet directions form the candidate direction set $V_p$, as shown in step~\textcircled{4} of Figure~\ref{fig:section3}(a). 

\subsection{Tail-Guided Sampling}\label{sec:tail-samp}
Under extreme selectivity, uniform sampling may miss rare-class documents. \aurora therefore favors both ends of the rough score distribution to improve class coverage and obtain more representative evidence for predicate-direction selection.


\smallskip\noindent\textbf{Rough Scoring}. 
\aurora first centers each pretrained embedding $\mathbf{x}_i$ using the corpus mean \(\mathbf{\mu}=\frac{1}{n}\sum_{j=1}^{n}\mathbf{x}_j\), writing \(\bar{\mathbf{x}}_i=\mathbf{x}_i-\mathbf{\mu}\). It then assigns each document a rough predicate score:
\begin{equation}
\label{eq:rough-score}
\begin{aligned}
  \ell^{\mathrm{rough}}_i=
  \max_{\mathbf{v}\in V_p}
  \phi(\bar{\mathbf{x}}_i)^\top \mathbf{v}.
\end{aligned}
\end{equation}
Eq.~\eqref{eq:rough-score} measures the maximum alignment between $d_i$ and the candidate directions. An informative direction assigns higher scores to positive documents and lower scores to negative ones. Step~\textcircled{5} of Figure~\ref{fig:section3}(b) presents the example score distribution of positive and negative documents.

\smallskip\noindent\textbf{Selectivity-adaptive Sampling}. 
The sampling bias depends on predicate selectivity: a balanced predicate requires near-uniform sampling, whereas an extremely selective predicate requires greater emphasis on the two tails.

Let $S_{\text{seed}}$ denote the indices of the five oracle-labeled documents used in Section~\ref{sec:pred-direction}. \aurora draws a \(1\%\) uniform sample $S_0$ from the remaining documents, \(D \setminus S_{\text{seed}}\), and obtains their oracle labels. It estimates predicate selectivity as \(\hat\theta=\frac{1}{|S_0|}\sum_{i\in S_0}y_i\). Let $\mathcal{R}(\ell_i^\mathrm{rough}) \in [0, 1]$ denote the normalized rank of $d_i$ in ascending order of its rough score. \aurora defines the tail factor as \(t_i=(2\left|\mathcal{R}(\ell^{\mathrm{rough}}_i)-\frac{1}{2}\right|)^3\). The tail factor is small near the middle of the ranking and large near its two ends. \aurora assigns each remaining document an unnormalized sampling weight:
\begin{equation}
\label{eq:tail-guided-pi}
\begin{aligned}
  w_i=
  \lambda(\hat\theta)
  +(1-\lambda(\hat\theta))t_i,
\end{aligned}
\end{equation}
where \(\lambda(\hat\theta)=4\hat\theta(1-\hat\theta)\). It then normalizes these weights:
\begin{equation}
\begin{aligned}
    p_i^\mathrm{tail}=\frac{w_i}{\sum_{j\notin S_{\text{seed}}\cup S_0}w_j},i\notin S_{\text{seed}}\cup S_0
\end{aligned}
\end{equation}
When $\hat{\theta}$ is close to $1/2$, $\lambda(\hat{\theta})$ approaches one and the distribution becomes nearly uniform. When $\hat{\theta}$ approaches zero or one, the distribution places more weight on the two tails. 

\aurora then draws a \textit{\(2\%\) tail-guided sample \(S_1\)} without replacement according to $p_i^\mathrm{tail}$. The accumulated labeled is \(S=S_{\text{seed}}\cup S_0 \cup S_1\).


\subsection{Label-Informed Direction Selection}\label{sec:di-select}
The candidate directions in $V_p$ are initial hypotheses and may not match the target collection. \aurora uses the labeled set \(S\) to retain discriminative directions and add a collection-specific direction.

\smallskip\noindent\textbf{Selecting Discriminative Candidates}. For each candidate direction $\mathbf{v}\in V_p$, \aurora computes a separation score:
\begin{equation}
\label{eq:sampled-signed-alignment}
  s^{\mathrm{sep}}_S(\mathbf{v})=
  \frac{1}{|S|}\sum_{i\in S}(2y_i-1)
  \phi(\bar{\mathbf{x}}_i)^\top \mathbf{v} .
\end{equation}
A positive document contributes positively when it aligns with \(\mathbf{v}.\) A negative document contributes positively when it points in the opposite direction. A larger score therefore indicates better separation between the two groups. \aurora ranks the candidates by this score and keeps the top $k_1$-1. We set $k_1$ = 12 by default. Directions out of the top $k_1$-1 are seen as bad and removed.


\myparagraph{Adding a Label-Derived Direction.} The retained directions encode LLM-generated predicate semantics but may miss collection-specific patterns. To capture these patterns, \aurora derives an additional direction from the oracle-labeled documents. Let \(S^+\) and \(S^-\) denote the index sets of positive and negative documents in \(S\), respectively. Their corresponding mean-centered embeddings are:
\begin{equation}
\label{eq:semantic-class-means}
  \mathbf{\mu}_S^+=\frac{1}{|S^+|}\sum_{i\in S^+}\bar{\mathbf{x}}_i,
  \qquad
  \mathbf{\mu}_S^-=\frac{1}{|S^-|}\sum_{i\in S^-}\bar{\mathbf{x}}_i.
\end{equation}
For each embedding dimension $j$, the within-class variance is:
\begin{equation}
\label{eq:semantic-fisher-var}
  \sigma_{S,j}^2=
  \frac{1}{2}{\operatorname{Var}}_{i\in S^+}(\bar{\mathbf{x}}_{ij})
  +
  \frac{1}{2}{\operatorname{Var}}_{i\in S^-}(\bar{\mathbf{x}}_{ij}).
\end{equation}
The label-derived direction is:
\begin{equation}
\label{eq:semantic-fisher}
  \mathbf{v}_f
  =
  \phi\!\left(
    (\mathbf{\mu}_S^+-\mathbf{\mu}_S^-)\oslash(\boldsymbol{\sigma}_S^2+\rho)
  \right),
\end{equation}
where \(\oslash\) denotes element-wise division and \(\boldsymbol{\sigma}_S^2=(\sigma_{S,1}^2,\ldots,\sigma_{S,m}^2)\). This construction follows a regularized diagonal Fisher form~\cite{fisher1936use}, denoted by $\mathbf{v}_f=\mathcal{F}(S,\{\bar{\mathbf{x}}_i\})$. It emphasizes dimensions with a large between-class mean and small within-class variation. 
\(\rho\) is set to a small value to prevent division by zero.

Finally, the \(k_1-1\) retained candidate directions with \(\mathbf{v}_f\) span a predicate-discriminative subspace.
We orthonormalize them to obtain the basis matrix \(\mathbf{C}_p\). 

\begin{algorithm}[t]\small
    \SetArgSty{textnormal}
    \SetAlgoLined
    \SetAlgoNoEnd
    \LinesNumbered
    \caption{Predicate-Aware Semantic Vector Generation}
    \KwIn{Centered document vectors \(\{\bar{\mathbf{x}}_i\}_{i=1}^{n}\), basis matrix \(\mathbf{C}_p\), overlap threshold \(\gamma\)}
    \KwOut{Semantic vector collection \(\{{\mathbf{z}}_i\}_{i=1}^{n}\)}
    \label{alg:semantic-vector-generation}

    Construct the document-vector-space matrix
    \(\bar{\mathbf{X}}\gets[\bar{\mathbf{x}}_1;\ldots;\bar{\mathbf{x}}_n]\)\; \label{alg:semvec:matrix}
    Compute the singular value decomposition
    \(\bar{\mathbf{X}}=\mathbf{U}\boldsymbol{\Sigma}\mathbf{V}^\top\), where rows of
    \(\mathbf{V}^\top\) are corpus-level principal directions\; \label{alg:semvec:svd}
    Initialize the nuisance basis \(\mathbf{N}_p\gets[\,]\)\; \label{alg:semvec:init-nuisance}
    \ForEach{principal direction \(\mathbf{q}_j\in\mathbf{V}^\top\)}{
        Compute overlap \(o_j\gets\norm{\mathbf{q}_j\mathbf{C}_p}\)\; \label{alg:semvec:overlap}
        \If{\(o_j < \gamma\) }{
            Append \(\mathbf{q}_j\) to \(\mathbf{N}_p\)\; \label{alg:semvec:append-nuisance}
        }
    }
    \For{\(i=1,\ldots,n\)}{
        Remove nuisance projection:
        \(\mathbf{r}_i\gets\bar{\mathbf{x}}_i-\bar{\mathbf{x}}_i\mathbf{N}_p\mathbf{N}_p^\top\)\; \label{alg:semvec:remove-nuisance}
        Project onto the predicate-discriminative subspace and normalize:
        \({\mathbf{z}}_i\gets\phi(\mathbf{r}_i\mathbf{C}_p)\)\; \label{alg:semvec:project}
    }
    \Return \(\{{\mathbf{z}}_i\}_{i=1}^{n}\)\;
\end{algorithm}

\subsection{Semantic Vector Construction}\label{sec:pred-vec-gen}
The predicate-discriminative subspace captures variation that separates the two classes. However, directly projecting document embeddings may still be dominated by high-variance corpus-level directions weakly related to the predicate. \aurora removes these nuisance directions before constructing the semantic vectors.


Algorithm~\ref{alg:semantic-vector-generation} shows the construction. \aurora first builds the centered
document matrix and applies singular value decomposition to obtain
corpus-level principal directions (Lines~\ref{alg:semvec:matrix}--\ref{alg:semvec:svd}). For each direction  \(\mathbf{q}_j\), it measures the overlap with the predicate-discriminative subspace as \(o_j=\norm{\mathbf{q}_j\mathbf{C}_p}\). 
Directions with \(o_j < \gamma\) form the nuisance basis \(\mathbf{N}_p\) 
(Lines~\ref{alg:semvec:init-nuisance}--\ref{alg:semvec:append-nuisance}). We set \(\gamma=0.1\) by default.
This threshold removes nearly orthogonal directions while retaining those with nontrivial predicate alignment. For each document, \aurora
removes its projection onto \(\mathbf{N}_p\) (Line~\ref{alg:semvec:remove-nuisance}) and projects the reminder onto \(\mathbf{C}_p\). The normalized result \({\mathbf{z}}_i\) is its predicate-aware semantic vector (Line~\ref{alg:semvec:project}). As shown in step~\textcircled{8} of Figure~\ref{fig:section3}(d), the semantic vector space provides a clearer separation between positive and negative documents than the original document vector space obtained in step~\textcircled{1} of Figure~\ref{fig:section3}(a).

\section{Lightweight Proxy with Progressive Refinement}\label{sec:light-proxy-model}

\begin{figure}[t]
  \centering
  \vspace{-1.25em}
  \includegraphics[width=0.48\textwidth]{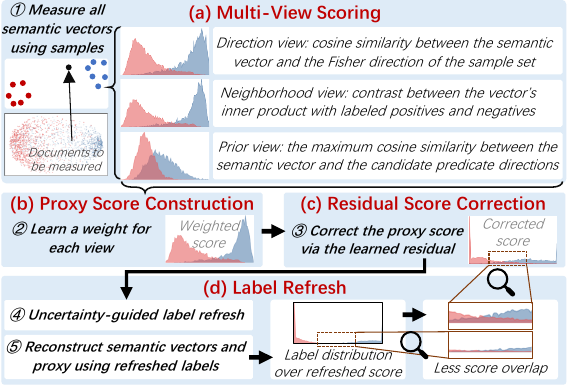}
  \vspace{-2.00em}
  \caption{Lightweight proxy construction and progressive refinement in \aurora.}
  \vspace{-1.5em}
  \label{fig:section4}
\end{figure}


A standard classifier may overfit the small labeled set. It also optimizes predictions under a single threshold, whereas a cascade requires reliable scores at both extremes. \aurora therefore constructs a lightweight proxy tailored to two-threshold routing.

To achieve this polarization, \aurora constructs a proxy score in four stages: 
1) It first scores each semantic vector from three complementary views (Figure~\ref{fig:section4} (a)) in  
Section~\ref{sec:multi-view scoring}. 2) It then normalizes and combines these view scores, learning their weights with an objective tailored to the target F1 requirement while polarizing the score distribution toward the two endpoints (Figure~\ref{fig:section4} (b)) in Section~\ref{sec:proxy-score-construction}. 3) \aurora further corrects systematic errors left by the linear mixture (Figure~\ref{fig:section4} (c)) in Section~\ref{sec:residual-correction}. 4) Finally, it acquires additional oracle labels from score ranges where positive and negative documents remain difficult to distinguish, and rebuilds both the semantic vectors and the proxy using the enlarged labeled set (Figure~\ref{fig:section4} (d)) in Section~\ref{sec:score-refresh}.  

\subsection{Multi-View Scoring}
\label{sec:multi-view scoring}
A single scoring signal is prone to instability on a small labeled set. \aurora therefore scores each semantic vector from three views: global class separation, local similarity to labeled documents, and alignment with the LLM-generated predicate descriptions.

\myparagraph{Direction View}
This view captures the global separation between labeled positive and negative documents. \aurora applies the regularized diagonal Fisher constructor \(\mathcal{F}\) from Section~\ref{sec:di-select} to the labeled semantic vectors:
\begin{equation}
\label{eq:fisher-score}
  \mathbf{h}_f=
  \mathcal{F}\!\left(S,\{{\mathbf{z}}_i\}\right),
  \qquad
  \ell_i^{\mathrm{dir}}={\mathbf{z}}_i^\top \mathbf{h}_f .
\end{equation}
The resulting score is larger when \(\mathbf{z}_i\) aligns with semantic dimensions that consistently distinguish labeled positives from negatives.


\myparagraph{Neighborhood View}
A single direction summarizes the global class separation but may miss local variation among documents. \aurora therefore also compares each document with nearby labeled examples from both classes. Let \(T^+_{i}\) denote the \(k_2\) labeled vectors in \(S^+\) with largest inner product with \({\mathbf{z}}_i\), and define \(T^-_{i}\) analogously for \(S^-\). We use \(k_2=5\) by default. Let \(\bar{\mathbf{z}}_{S}^{+}\) and \(\bar{\mathbf{z}}_{S}^{-}\) denote the positive and negative centroids of the labeled sample, respectively. For each document, \aurora compares:
\begin{equation}
\label{eq:exemplar-score}
\begin{aligned}
  \ell_i^{\mathrm{nbr}}
  =&
  \frac{1}{|T^+_{i}|}\sum_{\mathbf{u}\in T^+_{i}}{\mathbf{z}}_i^\top \mathbf{u}
  -\frac{1}{|T^-_{i}|}\sum_{\mathbf{u}\in T^-_{i}}{\mathbf{z}}_i^\top \mathbf{u}
  +\frac{1}{2}{\mathbf{z}}_i^\top(\bar{\mathbf{z}}_{S}^{+}-\bar{\mathbf{z}}_{S}^{-}).
\end{aligned}
\end{equation}
The first two terms indicate whether \(\mathbf{z}_i\) is more similar to nearby labeled positives than to nearby labeled negatives. The final term stabilizes this local comparison using the global difference between the two labeled centroids.

\myparagraph{Semantic-Prior View}
The third view reuses the rough predicate score derived from the LLM-generated anchors and facets in Eq.~\eqref{eq:rough-score}:
\begin{equation}
\label{eq:prior-score-view}
  \ell_i^{\mathrm{prior}}=\ell^{\mathrm{rough}}_i .
\end{equation}
Unlike the other views, the prior view does not require labels from both classes and remains available when \(S\) contains only one class.

\subsection{F1-Aligned View Combination}
\label{sec:proxy-score-construction}
\noindent\textbf{Multi-view Mixture}. Since the three views have different scales, \aurora converts their scores into normalized ranks before combination. Let \(U=\{\mathrm{dir},\mathrm{nbr},\mathrm{prior}\}\). For each view, \aurora maps its scores to normalized ranks \(r_i^u=\mathcal{R}(\ell_i^u)\), 
and forms the vector \(\mathbf{r}_i=[r_i^\mathrm{dir},r_i^\mathrm{nbr},r_i^\mathrm{prior}]^\top\).
It then combines views into a single score using a convex weight vector, \(\mathbf{w}\in\Delta_U\), where \(\Delta_U=\{\mathbf{w}:\mathbf{w}_u\ge0,\sum_{u\in U}\mathbf{w}_u=1\}\). The resulting multi-view score is:
\begin{equation}
\label{eq:proxy-mixture}
  \ell_i^\mathrm{mv}(\mathbf{w})=\mathbf{w}^{\top}\mathbf{r}_i .
\end{equation}
\aurora learns only three view weights instead of fitting a classifier directly on the high-dimensional semantic vectors, which improves stability when \(S\) is small.

\aurora initializes \(\mathbf{w}\) from the average rank gap between \(S^+\) and \(S^-\) on each view and assigns more weight to views with better sample separation.
When a class is absent (\(|S^+||S^-|=0\)) or no view shows a positive gap (\(\sum_{u\in U}[\Delta_v]_+=0\), where \([\Delta_u]_+=\max(\Delta_u,0)\)), it falls back to equal weights:
\begin{equation}
\label{eq:proxy-init}
\begin{aligned}
  \Delta_u
  &=
  \begin{cases}
  \frac{1}{|S^+|}\sum_{i\in S^+}r_i^u
  -
  \frac{1}{|S^-|}\sum_{i\in S^-}r_i^u,
  & |S^+||S^-|>0,\\
  0, & \text{otherwise},
  \end{cases}\\
  \mathbf{w}^{(0)}_u
  &=
  \begin{cases}
  \frac{[\Delta_u]_+}{\sum_{v\in U}[\Delta_v]_+},
  & \sum_{v\in U}[\Delta_v]_+>0,\\
  \frac{1}{|U|}, & \text{otherwise},
  \end{cases}
\end{aligned}
\end{equation}

\myparagraph{Additive F1 Budget}
The view weights should favor score distributions that support the target F1. Directly optimizing F1 is difficult because it is a nonlinear ratio of collection-level counts:
\begin{equation}
\label{eq:f1-definition}
  \text{F1}=
  \frac{2\sum_{i=1}^{n}\hat y_i y_i}
       {2\sum_{i=1}^{n}\hat y_i y_i
       +\sum_{i=1}^{n}\hat y_i(1-y_i)
       +\sum_{i=1}^{n}(1-\hat y_i)y_i}.
\end{equation}
\aurora therefore rewrites the constraint \(\text{F1}\geq\alpha\) as an equivalent additive document-level budget. For a threshold pair \(\tau=(l,r)\), define the per-document budget contribution:
\begin{equation}\label{eq:f1-budget}
  \begin{aligned}
  \mathcal{Z}_{i}(\alpha,\tau;\mathbf{w})
  =
  2(1-\alpha)y_i
  -(2-\alpha)L_i(\tau)y_i
  -\alpha H_i(\tau)(1-y_i),
  \end{aligned}
\end{equation}
where \(L_i(\tau)=\mathbb{I}[\ell^\mathrm{mv}_i(\textbf{w})\le l]\) and
\(H_i(\tau)=\mathbb{I}[\ell_i^\mathrm{mv}(\textbf{w})\ge r]\) are the indicators. Summing over documents gives the F1 budget $\mathcal{B}_{\alpha}(\tau;\mathbf{w})=\sum_{i\in D}\mathcal{Z}_{i}(\alpha,\tau;\mathbf{w})$, and then $\text{F1}\geq\alpha \Longleftrightarrow \mathcal{B}_{\alpha}(\tau)\ge0$.
A positive document, whether resolved by the proxy or the oracle, contributes $2(1-\alpha)$. A false positive or false negative contributes $-\alpha$, while a correctly labeled negative contributes zero. 
We defer the algebra of this equivalence to {Appendix~\ref{app:f1-budget}}.


\myparagraph{F1-Aligned Weight Learning} The oracle labels required to compute \(\mathcal{Z}_i\) are available only for documents in \(S\). For the unlabeled documents, \aurora uses its multi-view score as a soft surrogate for the unknown label, i.e., \(\tilde{y}_i(\mathbf{w})=\ell_i^\mathrm{mv}(\mathbf{w})\) and defines a surrogate budget contribution:
\begin{equation}
\label{eq:f1-budget-pair}
\begin{aligned}
  \widetilde{\mathcal{Z}}_i(\alpha,\tau;\mathbf{w})
  =
  2(1-\alpha)\tilde{y}_i(\mathbf{w})
  -(2-\alpha)L_i(\tau)\tilde{y}_i(\mathbf{w})
  -\alpha H_i(\tau)(1-\tilde{y}_i(\mathbf{w})).
\end{aligned}
\end{equation}

Optimizing for one threshold pair may overfit the score to that pair. \aurora therefore samples a set \(A\) of candidate pairs \(\tau=(l,r)\) from \([0,1]^2\)  with \(l < r\), and minimizes the average variance of the budget residual over \(A\). This encourages more stable F1 budget estimates, which in turn helps produce tighter confidence radius during threshold selection in Section~\ref{sec:th-select-wg}.
Minimizing this variance alone can overfit the small labeled sample. \aurora therefore adds two penalties, giving the final objective:
\begin{equation}
\label{eq:polar-objective}
\begin{aligned}
  \mathbf{w}^*
  =
  &\arg\min_{\mathbf{w}\in\Delta_U}
  \frac{1}{|A|}\sum_{\tau\in A}
  {\operatorname{Var}}_{i\in S}
  \!\left[
    \mathcal{Z}_{i}(\alpha,\tau;\mathbf{w})
    -\widetilde{\mathcal{Z}}_i(\alpha,\tau;\mathbf{w})
  \right]\\
  &+\lambda_{1}\frac{1}{n}\sum_{i=1}^{n}\ell_i^\mathrm{mv}(\mathbf{w})(1-\ell_i^\mathrm{mv}(\mathbf{w}))\\
  &+\lambda_{2}\left[\frac{1}{|S^+|}\sum_{i\in S^+}(\ell_i^\mathrm{mv}(\mathbf{w})-1)^2+\frac{1}{|S^-|}\sum_{i\in S^-}\ell_i^\mathrm{mv}(\mathbf{w})^2\right].
\end{aligned}
\end{equation}
{The first term stabilizes the F1 budget with the observed budget across different threshold pairs. The second term pushes collection-wide scores away from the middle of the score range, while the third pulls labeled positives and negatives toward $1$ and $0$, respectively. 
The latter two terms increase the number of documents that the cascade can resolve without the oracle. \aurora optimizes the weights using Adam \cite{kingma2015adam}.


\subsection{Residual Score Correction}
\label{sec:residual-correction}
A linear combination of the three scores may remain insufficient to distinguish both classes. \aurora therefore learns a lightweight nonlinear residual model on \(S\) to correct the residual scoring errors.


\myparagraph{Residual Features} The residual model should be more expressive than reweighting the three views, while remaining stable under a small labeled set. Let \(\bar{\mathbf{z}}_{S}^{+}\) and \(\bar{\mathbf{z}}_{S}^{-}\) denote the positive and negative centroids of the labeled sample, respectively. When both classes are present, \aurora computes:
\begin{equation}
\label{eq:corr-centroids}
\begin{aligned}
  \beta_i
  &= \mathcal{R}\!\left(
  {\mathbf{z}}_i^\top\phi(\bar{\mathbf{z}}_{S}^{+})
  -{\mathbf{z}}_i^\top\phi(\bar{\mathbf{z}}_{S}^{-})
  \right)-\frac{1}{2}.
\end{aligned}
\end{equation}
When $S$ contains only one class, \aurora sets \(\beta_i=0\). A larger \(\beta_i\) indicates that $d_i$ is more strongly aligned with the labeled positive centroid than with the negative centroid. The final feature vector is
\begin{equation}
\label{eq:corr-feature}
  \mathbf{f}_i=
  \left[
    {\mathbf{z}}_i^\top,
    \ell_i^\mathrm{mv}(\mathbf{w}^{*}),
    \left(\ell_i^\mathrm{mv}(\mathbf{w}^{*})\right)^2,
    \beta_i,\ 1
  \right]^\top .
\end{equation}
The semantic vector \(\mathbf{z}_i\) retains predicate-specific information that may be lost in the three view scores. The linear and quadratic score terms capture calibration errors, while \(\beta_i\) provides a class-relative margin. Together, these features capture residual patterns that view reweighting cannot correct.

\myparagraph{Residual Model} \aurora fits a ridge regression model to predict the residual between the oracle label and the multi-view score:
\begin{equation}
\label{eq:ridge-residual}
  \boldsymbol{\gamma}^*=
  \arg\min_{\boldsymbol{\gamma}}
  \sum_{i\in S}\left(
  y_i-\ell_i^\mathrm{mv}(\mathbf{w}^{*})-\boldsymbol{\gamma}^\top\mathbf{f}_i
  \right)^2
  +\lambda_{3}\norm{\boldsymbol{\gamma}}^2 .
\end{equation}
The ridge penalty controls overfitting to the small labeled set. 
For each document, the learned residual correction is \(c_i=\boldsymbol{\gamma}^{\top}\mathbf{f}_i\). We set \(\lambda_3=2.0\) by default.

\myparagraph{Score Stretching} \aurora first adds the learned correction to the multi-view score and normalizes the values \(\ell^\mathrm{res}_i=\mathcal{R}(\ell_i^\mathrm{mv}(\mathbf{w}^{*})+c_i)\). 
\aurora then applies an order-preserving stretching function:
\begin{equation}
\label{eq:corr-score}
  \begin{aligned}
  \ell^\mathrm{corr}_i
  &= 
  \frac{({\ell}^\mathrm{res}_i)^2}
       {({\ell}^\mathrm{res}_i)^2+(1-{\ell}^\mathrm{res}_i)^2}.
  \end{aligned}
\end{equation}
The transformation keeps $0.5$ unchanged while pushing scores below and above $0.5$ toward $0$ and $1$, respectively. 
The transformation preserves score order. 
Section~\ref{sec:score-refresh} uses the stretched scores to emphasize documents that remain near $0.5$.
\subsection{Iterative Refinement via Label Refresh}
\label{sec:score-refresh}
Tail-guided sampling improves class coverage but may leave overlapping score regions insufficiently explored. Inspired by uncertainty sampling in active learning~\cite{lewis1994sequential,settles2009active}, \aurora prioritizes these poorly resolved regions for additional oracle labeling to refine the proxy's discrimination. It performs two refinement rounds, each labeling another 2\% of the collection.

\myparagraph{Uncertainty-guided Label Refresh} 
For each unlabeled document \(d_i\) with score \(\ell_i^\mathrm{corr}\), \aurora computes an uncertainty weight \(a_i=\sqrt{\ell_i^\mathrm{corr}(1-\ell_i^\mathrm{corr})}+\epsilon\), where \(\epsilon=0.02\) keeps every document eligible for sampling. The uncertainty is largest when \(\ell_i^\mathrm{corr}\) is near \(0.5\) and decreases toward the score ends. 

Sampling only by uncertainty may concentrate labels in a narrow score range. 
\aurora therefore partitions the unlabeled documents into equal-frequency score bins and uncertainty bins.
Each nonempty intersection of a score bin and an uncertainty bin forms a stratum \(G\). 

Given a per-round refresh budget \(|S_{\text{ref}}|=0.02n\), \aurora allocates 
\begin{equation}
\label{eq:residual-neyman}
  n_G
  =
  \left\lfloor
    |S_{\text{ref}}|\,
    \frac{|G|\bar a_G}{\sum_{G'}|G'|\bar a_{G'}}
  \right\rfloor,
  \qquad
  \bar a_G=\frac{1}{|G|}\sum_{i\in G}a_i ,
\end{equation}
labels to stratum \(G\). This Neyman-inspired allocation favors large, uncertain strata while preserving coverage across the score range.

\myparagraph{Iterative Refitting} The newly labeled documents in \(S_{\text{ref}}\) are added to \(S\) and \aurora reruns the construction from Section~\ref{sec:di-select} through Section~\ref{sec:residual-correction} using new $S$. 
The updated corrected proxy scores then guide the next refresh round. Figure~\ref{fig:section4}(d) shows the document distribution before and after label refresh. \aurora progressively improves score separation in poorly resolved score ranges while maintaining coverage of the full score distribution.



\section{Quality-Certified Threshold Selection}\label{sec:th-select-wg}
Given the corrected proxy scores, \aurora selects a threshold pair  \(\tau=(l,r)\) that maximizes oracle-call reduction while satisfying the target F1. The reduction of each pair is known from the scores, whereas its F1 depends on the unknown labels of proxy-resolved documents. Directly estimating F1 from a small, non-uniform sample is difficult because F1 is a nonlinear ratio of collection-level counts. \aurora uses the additive F1 budget $\mathcal{B}_{\alpha}$ as Section~\ref{sec:proxy-score-construction}.


A naive approach would estimate \(\mathcal{B}_{\alpha}(\tau)\) for every possible \(\tau\) and select the \(\tau\) with the largest reduction and \(\mathcal{B}_{\alpha}(\tau)\ge0\). This approach is both expensive and unsafe: the number of \(\tau\)s is large, and sampling error may make an infeasible pair appear to satisfy the F1 target over the full collection. \aurora addresses these challenges in four steps. It first obtains an audit sample designed for threshold-quality estimation (Section~\ref{sec:audit-sample}). It then estimates the quality and reduction of candidate threshold pairs and retains a compact candidate set (Section~\ref{sec:PF-gen}). Next, \aurora uses the candidates to tighten the budget estimates and construct simultaneous lower confidence bounds (Section~\ref{sec:th-select}). Finally, it selects the certified \(\tau\) with the largest oracle-call reduction and establishes the quality guarantee (Section~\ref{sec:quality-guan}).

\subsection{Score-Bin Pilot Sampling}\label{sec:audit-sample}
The labels in \(S\) were acquired under intermediate proxy scores for proxy construction and may not adequately represent the final corrected-score distribution. Directly using \(S\) may therefore yield an inaccurate or high-variance estimate of the collection-level F1 budget. Falcon therefore draws a separate audit sample \(S_a\) from \(D\setminus S\). \aurora uses a total audit budget of \(|S_a|=0.05\times|D|\) by default and allocates it in two stages. We introduce the first stage here.

\myparagraph{Score-bin Pilot Sampling}
\aurora first partitions the documents in \(D\setminus S\) according to their \(\ell_i^\mathrm{corr}\). It divides $[0,1]$ into intervals of width $0.01$, merges the empty bins with nonempty neighbors, and generates the score-bin collection \({B}^{\text{score}}\).
Documents within the same bin have close scores and are therefore likely to receive the same proxy decision under nearby threshold pairs.

For each score bin \(b\in {B}^{\text{score}}\), \aurora uniformly selects \(m^{(0)}_b=\lceil 1\%\cdot |b|\rceil\) documents, \(S^{(0)}_{a,b}\), and uses the sample to estimate the positive document rate of \(b\) as \(\hat{p}_b=\frac{\sum_{i\in S_b}{y_i}}{|m^{(0)}_b|}\). \aurora then uses \(q_{b}=\sqrt{\widehat p_{b}(1-\widehat p_{b})}\) to measure the estimated label variation of \(b\).

\subsection{Pareto Frontier Generation}\label{sec:PF-gen}
A threshold pair with a larger oracle-call reduction resolves more documents without the oracle, but may also introduce more false positives or false negatives. Candidate threshold pairs therefore exhibit a trade-off between oracle-call reduction and the F1 budget. \aurora uses the collected samples to estimate this trade-off and compress the large threshold space into a compact Pareto frontier.

\aurora constructs an initial candidate pool, $\mathcal{T}_0$, by randomly drawing \(100{,}000\) threshold pairs $(l,r)$ with \(l<r\), where \(l\) and \(r\) come from the sorted corrected scores $\{\ell^\mathrm{corr}_i\}_{i=1}^n$. For each \(\tau \in \mathcal{T}_0\), \aurora computes its oracle-call reduction exactly and estimates its collection-level F1 budget from the available oracle labels. 

\myparagraph{Metrics Estimation}
For a candidate threshold pair \(\tau=(l,r)\), we define the {oracle reduction rate} used to construct the frontier as:
\begin{equation}
\label{eq:frontier-reduction}
   \mathrm{Red}(\tau)=
  \frac{
    |\{i\notin S \cup S_a^{(0)}:
    \ell^\mathrm{corr}_i\leq l\ \mathrm{or}\ \ell^\mathrm{corr}_i \geq r\}|
  }{|D|}.
\end{equation}

Denote \(S_a^{(0)}\) as the pilot sample set obtained in Section~\ref{sec:audit-sample}.
For each \(\tau \in \mathcal{T}_0\), \aurora estimates the collection-level F1 budget on the labeled set \(S\cup S_a^{(0)}\) as:
\begin{equation}
\label{eq:frontier-budget-signal}
\widehat{\mathcal{B}}_{\alpha}(\tau)
=
\sum_{i \in S}\mathcal{Z}_i(\alpha,\tau)
+
\sum_{b \in {B}^{\mathrm{score}}}
\frac{|b|}{m_b^{(0)}}
\sum_{i \in S_{a,b}^{(0)}}\mathcal{Z}_i(\alpha,\tau).
\end{equation}
\(\mathcal{Z}_i(\alpha,\tau)\) is the F1 budget contribution of $d_i$, defined in Section~\ref{sec:proxy-score-construction} and we omit \(\mathbf{w}\) here. The first term is the contribution of documents in \(S\). In each score bin, \aurora expands the pilot sample by the inverse sampling fraction to estimate the contribution of the remaining documents in the same bin. Summing over all bins yields the estimated collection-level F1 budget. Appendix~\ref{app:budget-estimator} proves that the estimator is unbiased.


\myparagraph{Pareto Filtering}
With the \((\widehat{\mathcal{B}}_{\alpha}(\tau),\text{Red}(\tau))\) of each $\tau$, \aurora then compresses $\mathcal{T}_0$ by Pareto filtering. A pair \(\tau_1\) is discarded if there exists any other pair \(\tau_2\) that has $\widehat {\mathcal{B}}_\alpha(\tau_2) \ge \widehat {\mathcal{B}}_\alpha(\tau_1)$ and \(\mathrm{Red}(\tau_2) \geq \mathrm{Red}(\tau_1)\). \aurora additionally retains the candidates with the largest reduction. These representatives, together with the non-dominated candidates, form the candidate frontier \(\mathcal{T}\). 

\subsection{Threshold Selection}\label{sec:th-select}
The estimate \(\widehat {\mathcal{B}}_\alpha(\tau)\) remains subject to sampling error. Directly selecting the largest-reduction candidate with a nonnegative estimated budget may therefore accept a threshold pair whose true collection-level budget is negative. 

\aurora instead certifies each candidate using a lower confidence bound on its F1 budget.
To tighten this bound, \aurora uses the provisional frontier and pilot audit labels to reorganize the documents into frontier-aware strata. The score bins \({B}^{\text{score}}\) capture where a document lies on the corrected-score axis, while the candidate frontier reveals how that document is treated across different threshold pairs. Grouping documents with similar score and assignment behavior reduces the within-stratum variation of their F1-budget contributions and thereby tightens the resulting confidence bounds. 
\aurora then uses the remaining audit budget for post-frontier certification sampling to improve the final budget and variance estimates.


\setlength{\textfloatsep}{8pt plus 1pt minus 2pt}
\begin{algorithm}[t]\small
    \SetArgSty{textnormal}
    \SetAlgoLined
    \SetAlgoNoEnd
    \LinesNumbered
    \caption{Frontier-Aware Strata Construction}
    \KwIn{Candidate frontier \(\mathcal{T}\), corrected scores
    \(\{\ell_i^\mathrm{corr}\}_{i=1}^{n}\), audit score bins \({B}^{\text{score}}\), proxy sample \(S\),
    audit set \(S_a\)}
    \KwOut{Strata \(\mathcal{G}=\{G_1,\ldots,G_{|\mathcal{G}|}\}\) over
    \(D\setminus S\)}
    \label{alg:frontier-strata}

    Partition \([0,1]\) into assignment bins \({B}^\mathrm{assign}\) with width \(0.01\)\; \label{alg:frontier-strata:covbins}
    Initialize \(\mathcal{G}\) indexed by
    \(\{\mathrm{low},\mathrm{high},\mathrm{both},\mathrm{mid}\}
    \times {B}^{\text{score}} \times {B}^\mathrm{assign}\)\; \label{alg:frontier-strata:init-map}
    \ForEach{\(i\in D\setminus S\)}{
        \(\kappa_i^{-}\gets\sum_{\tau=(l,r)\in\mathcal{T}}\mathbb{I}[\ell^{corr}_i\le l]\), \(\kappa_i^{+}\gets\sum_{\tau=(l,r)\in\mathcal{T}}\mathbb{I}[\ell^{corr}_i\ge r]\)\; \label{alg:frontier-strata:low-coverage}
        \uIf{\(\kappa_i^{-}>0\) and \(\kappa_i^{+}=0\)}{
            \(\eta_i\gets\mathrm{low}\)\; \label{alg:frontier-strata:side-start}
        }
        \uElseIf{\(\kappa_i^{-}=0\) and \(\kappa_i^{+}>0\)}{
            \(\eta_i\gets\mathrm{high}\)\;
        }
        \uElseIf{\(\kappa_i^{-}>0\) and \(\kappa_i^{+}>0\)}{
            \(\eta_i\gets\mathrm{both}\)\;
        }
        \Else{
            \(\eta_i\gets\mathrm{mid}\)\; \label{alg:frontier-strata:side-end}
        }
        Find the score bin \(b_i^\mathrm{score}\in {B}^{\text{score}}\) that contains
        \(\ell_i^\mathrm{corr}\)\; \label{alg:frontier-strata:score-bin}
        \(u_i\gets(\kappa_i^{-}+\kappa_i^{+})/|\mathcal{T}|\)\; \label{alg:frontier-strata:coverage-ratio}
        Find the assignment bin \(b_i^\mathrm{assign}\in {B}^\mathrm{assign}\) that contains
        \(u_i\)\; \label{alg:frontier-strata:coverage-bin}
        Append \(d_i\) to \(\mathcal{G}[\eta_i,b_i^\mathrm{score},b_i^\mathrm{assign}]\)\; \label{alg:frontier-strata:append}
    }
    \ForEach{group index \(idx=(\eta_i,b_i^\mathrm{score},b_i^\mathrm{assign})\)}{
        Merge \(\mathcal{G}[idx]\) into the nearest group if \(|\mathcal{G}[idx]\cap S_a^{(0)}|<2\)\;\label{alg:frontier-strata:merge}
    }
    \Return the nonempty merged groups as
    \(\mathcal{G}=\{G_1,\ldots,G_{|\mathcal{G}|}\}\)\; \label{alg:frontier-strata:end}
\end{algorithm}

\myparagraph{Frontier-aware Strata}
Algorithm~\ref{alg:frontier-strata} presents strata construction over \(D\setminus S\). \aurora reuses the score bins \({B}^{\text{score}}\) and partitions \([0,1]\) into intervals of width \(0.01\) as the assignment bins \({B}^{\text{assign}}\) (Line~\ref{alg:frontier-strata:covbins}). For $d_i\in D\setminus S$, \aurora counts how many candidate threshold pairs automatically resolve it as negative or positive (Lines~\ref{alg:frontier-strata:low-coverage}). Based on the counts, \aurora assigns $d_i$ one of the four assignment categories, \(\eta_i\) (Lines \ref{alg:frontier-strata:side-start}-\ref{alg:frontier-strata:side-end}). The categories distinguish documents that may be automatically resolved only as negative, only as positive, on either side under different candidates, or never by any candidate. \aurora also computes the fraction of candidates, \(u_i\) that automatically resolve $d_i$, finds the score bin \(b_i^{\text{score}} \in {B}^{\text{score}}\) containing \(\ell_{i}^\mathrm{corr}\) and the assignment bin \(b_i^{\text{assign}}\) containing \(u_i\), and groups documents using the key \((\eta_i, b_i^{\text{score}},b_i^{\text{assign}})\) (Lines \ref{alg:frontier-strata:score-bin}-\ref{alg:frontier-strata:append}). Finally, \aurora merges any group containing fewer than two audit documents with its nearest group, resulting in the stratum collection, \(\mathcal{G}\) (Lines \ref{alg:frontier-strata:merge}-\ref{alg:frontier-strata:end}).



\myparagraph{Post-frontier Certification Sampling}
After candidate frontier and frontier-aware strata have been fixed, \aurora allocates remaining audit budget to the non-pilot documents in each retained stratum. Let \(\widehat{V}_{G_i}(\alpha)\) be the pilot-based estimate of budget-contribution variance in stratum \(G_i\). The certification allocation is:
\begin{equation}
\label{eq:audit-neyman}
  m^{(1)}_{G_i}
  =
  \left\lfloor
  \left(|S_a|-\sum_{b\in {B}^{\text{score}}}m_b^{(0)}\right)
  \frac{|G_i\setminus S_a^{(0)}|\sqrt{\widehat{V}_{G_i}(\alpha)}}
  {\sum_{G_j\in\mathcal{G}}|G_j\setminus S_a^{(0)}|\sqrt{\widehat{V}_{G_j}(\alpha)}}
  \right\rfloor.
\end{equation}
The resulting certification sample is \(S^{(1)}_{a,G_i}\), and the complete audit set is \(S_a=S_a^{(0)}\cup(\bigcup_{G_i\in\mathcal{G}}S^{(1)}_{a,G_i})\).


\myparagraph{Stratified Budget Estimator}
For a candidate \(\tau\in\mathcal{T}\), the revised mean F1 budget contribution in stratum \(G_i \in \mathcal{G}\) is:
\begin{equation}
\label{eq:stratified-budget-contribution}
  \widebar{\mathcal{Z}}_{G_i}(\alpha,\tau)
  =
  \frac{1}{|S_a\cap G_i|}
  \sum_{j\in S_a\cap G_i}\mathcal{Z}_j(\alpha,\tau).
\end{equation}
\aurora revises the estimated F1 budget by combining the known labels in \(S\) with the
revised estimate over \(D\setminus S\):
\begin{equation}
\label{eq:stratified-budget-estimator}
  \widehat{\mathcal{B}}_\alpha(\tau)
  =
  \sum_{i\in S}\mathcal{Z}_i(\alpha,\tau)
  +
\sum_{i=1}^{|\mathcal{G}|}|G_i|\widebar{\mathcal{Z}}_{G_i}(\alpha,\tau).
\end{equation}
For \(G_i\), the \emph{within-stratum sample variance} is:
\begin{equation}
\label{eq:stratified-budget-sample-var}
  s_{G_i}^2(\alpha,\tau)
  =
  \frac{1}{|S_a\cap G_i|-1}
  \sum_{j\in S_a\cap G_i}
  \left(\mathcal{Z}_j(\alpha,\tau)-\widebar{\mathcal{Z}}_{G_i}(\alpha,\tau)\right)^2.
\end{equation}
Because audit sampling is performed without replacement, \aurora uses the \emph{finite population corrected variance} estimator to quantify the uncertainty of the stratified budget estimate:
\begin{equation}
\label{eq:stratified-fpc-var}
  {\operatorname{Var}}\!\left[\widehat{\mathcal{B}}_\alpha(\tau)\right]
  =
  \sum_{i=1}^{|\mathcal{G}|}
  |G_i|^2
  \left(1-\frac{|S_a\cap G_i|}{|G_i|}\right)
  \frac{s_{G_i}^2(\alpha,\tau)}{|S_a\cap G_i|}.
\end{equation}
The factor \(1-|S_a\cap G_i|/|G_i|\) is the finite population correction.
Inspired by empirical Bernstein-style confidence bounds
\cite{serfling1974probability,bardenet2015concentration,maurer2009empirical},
\aurora defines the variance-calibrated radius as:
\begin{equation}
\label{eq:stratified-radius}
\begin{aligned}
  \operatorname{Rad}_\alpha(\tau)
  &=
  \sqrt{
    2{\operatorname{Var}}\!\left[\widehat{\mathcal{B}}_\alpha(\tau)\right]
    \log\!\left(\frac{3|\mathcal{T}|}{\delta}\right)
  },
\end{aligned}
\end{equation}
The lower confidence bound is \(\operatorname{LCB}_\alpha(\tau)=\widehat{\mathcal{B}}_\alpha(\tau)
  -
\operatorname{Rad}_\alpha(\tau)\).

\myparagraph{Final Selection}
The final threshold pair is the certified candidate with the largest
oracle-call reduction \(\mathrm{Red}(\tau)\) from Eq.~\eqref{eq:frontier-reduction}:
\begin{equation}
\label{eq:certified-final-selector}
  \tau^*
  =
  \arg\max_{\tau\in\mathcal{T}}\mathrm{Red}(\tau)
  \
  \mathrm{s.t.}\
  \operatorname{LCB}_\alpha(\tau)
  \ge0 .
\end{equation}
If no candidate satisfies the lower-bound constraint, \aurora routes all documents to the oracle.

\subsection{Quality Guarantee}\label{sec:quality-guan}
The following lemma establishes the simultaneous coverage of the lower confidence bounds constructed in Section~\ref{sec:th-select}.
\begin{lemma}[Simultaneous lower confidence bound]
\label{lem:simultaneous-lcb}
The simultaneous LCBs are asymptotically valid:
\begin{equation}
\label{eq:simultaneous-event}
  \liminf_{|D|\to\infty}\mathbb{P}\!\left[
    \mathcal{B}_\alpha(\tau)
    \geq
    \operatorname{LCB}_\alpha(\tau),
    \ \forall\tau\in\mathcal{T}
  \right]
  \geq
  1-\delta.
\end{equation}
\end{lemma}
The proof is provided in Appendix~\ref{app:proof-lcb}. Based on this result, we establish the quality guarantee of the threshold-selection procedure.


\begin{theorem}[Certified threshold selection under the calibrated bound]
\label{thm:certified-threshold}
The labeling returned by \aurora satisfies the target F1 constraint, \(\mathbb{P}\!\left[\mathrm{F1}\ge\alpha\right]\ge1-\delta\), asymptotically.
\end{theorem}

\begin{proof}
By Lemma~\ref{lem:simultaneous-lcb}, as \(|D|\to\infty\), \(\mathcal{B}_{\alpha}(\tau)\geq\operatorname{LCB}_{\alpha}(\tau)\)
holds for all \(\tau\in\mathcal{T}\) with asymptotic probability
at least \(1-\delta\). Let \(\tau^*\) denote the selected pair from Section~\ref{sec:th-select}. Then,
\(
  \mathcal{B}_\alpha(\tau^*)
  \ge
  \operatorname{LCB}_\alpha(\tau^*)
  \ge0,
\)
with probability at least \(1-\delta\). By the equivalence established in Section~\ref{sec:proxy-score-construction}, \(\text{F1}\geq\alpha \Longleftrightarrow \mathcal{B}_{\alpha}(\tau)\ge0\), this implies that the resulting labeling satisfies \(\text{F1}(\tau^*)\ge\alpha\) with probability at least \(1-\delta\).
\end{proof}

\begin{figure*}[t]
  \centering
  \includegraphics[width=\textwidth]{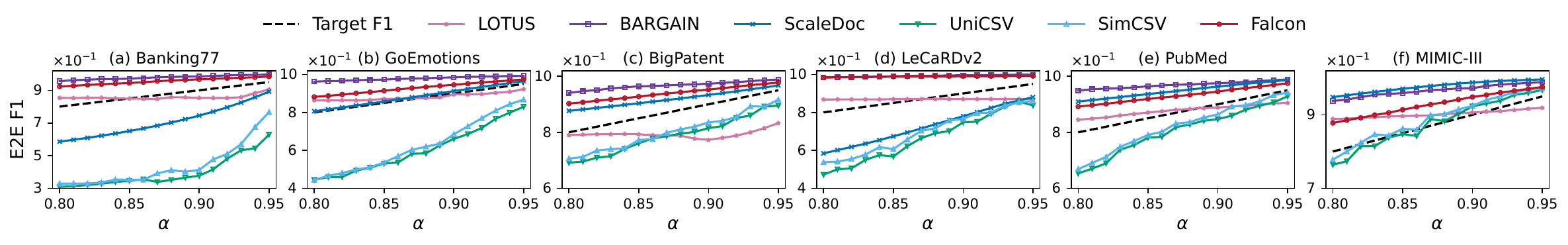}
  \vspace{-2.9em}
  \caption{Average achieved end-to-end F1 under varying target F1 constraints (5 random seeds per query).}
  \vspace{-1.35em}
  \label{fig:f1}
\end{figure*}
\section{Experiment} \label{sec:experiment}
\subsection{Experimental Setup} \label{sec:exp-setup}
\myparagraph{Datasets}
We select six real-world text collections spanning a wide range of document lengths and domains. Each dataset contains 20 semantic-filtering
queries with predicate selectivities ranging from \(0.65\%\)
to \(97.34\%\). Table~\ref{tab:dataset} summarizes the dataset statistics.

\begin{table}[t] \small
    \centering
    \caption{Dataset statistics.}
    \vspace{-1.00em}
    \label{tab:dataset}
    \setlength{\tabcolsep}{4.4pt}
    \scalebox{0.86}{\begin{tabular}{ccccc}
    \toprule
    \textbf{Datasets} & $|D|$ & \textbf{Avg. Chars.} &
    \textbf{Selectivity} & \textbf{Description} \\
    \midrule
    \bank \cite{casanueva2020banking77} & 10,000 & 59 & 1.20\%--83.68\% & Banking intents \\
    \emotion \cite{demszky2020goemotions} & 10,000 & 66 & 4.63\%--50.77\% & Social-media statements \\
    \patent \cite{sharma2019bigpatent} & 10,035 & 662 & 2.17\%--97.34\% & Patent abstracts \\
    \law \cite{li2024lecardv2} & 2,000 & 1,615 & 0.65\%--9.20\% & Legal case records \\
    \med \cite{dernoncourt2017pubmedrct} & 10,036 & 1,765 & 3.46\%--91.59\% & Biomedical abstracts \\
    \mimic \cite{johnson2016mimiciii} & 10,000 & 5,128 & 7.49\%--85.64\% & Clinical notes \\
    \bottomrule
    \end{tabular}}
\end{table}


\myparagraph{Comparisons}
We list the methods for comparison as follows:
\begin{itemize}[left=0pt]
    \item \lotus \cite{Patel25lotus}: It uses a smaller LLM to generate proxy scores and applies a variant of \texttt{SUPG}~\cite{kang2020supg} to reduce oracle calls.
    \item \bargain \cite{Zeighami25bargain}: A state-of-the-art method that first calls a smaller LLM to label all documents with a confidence score. An expensive LLM resolves uncertain documents under statistical guarantees.
    \item \scaledoc \cite{zhang2026scaledoc}: A cascade that combines a contrastively trained proxy with heuristic threshold selection. Documents not resolved by the proxy are routed to the LLM oracle.
    \item \texttt{UniCSV}, \texttt{SimCSV} \cite{hou26csv}: Clustering-based proxies that group document embeddings and sample representative ones for LLM labeling. They estimate cluster quality via voting, recursively refine uncertain clusters, and route unresolved to the oracle. 
\end{itemize}

\begin{figure*}[t]
  \centering
  \includegraphics[width=\textwidth]{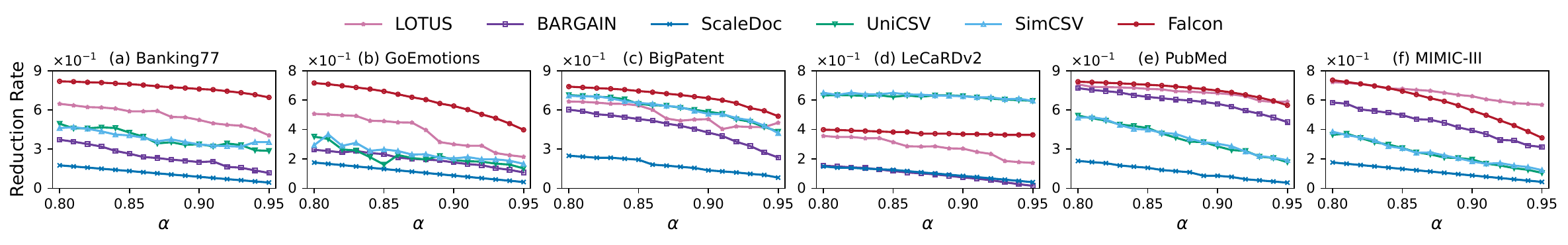}
  \vspace{-2.9em}
  \caption{Average oracle-call reduction under varying target F1 constraints (5 random seeds per query).
  }
  \vspace{-1.25em}
  \label{fig:pure_reduction}
\end{figure*}

\begin{figure}[t]
  \centering
  \includegraphics[width=0.485\textwidth]{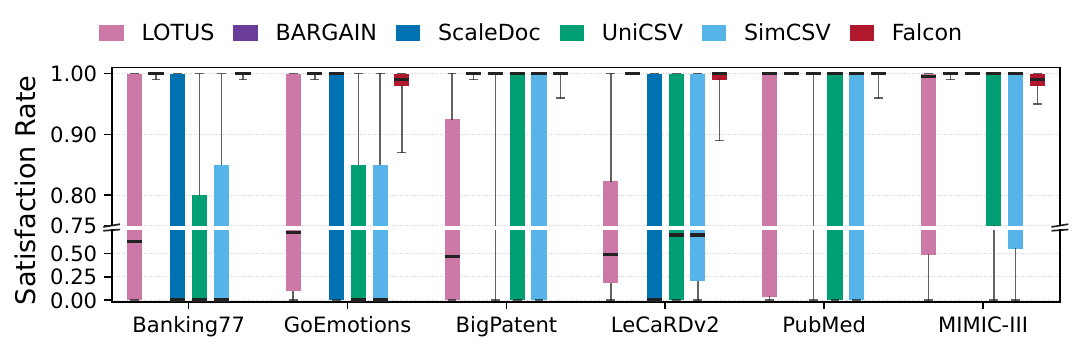}
  \vspace{-2.5em}
  \caption{Per-query quality satisfaction rate ($\alpha=0.9$). Box edges indicate the 25th and 75th percentiles.
  }
  \vspace{-0.75em}
  \label{fig:satisfaction}
\end{figure}

\myparagraph{Evaluation Aspects}
{For semantic filtering, \emph{high quality or reduction alone is not meaningful}: assigning arbitrary labels without LLM calls results in 100\% reduction, but arbitrarily poor quality; using the oracle for every document could achieve high quality, but 0\% reduction. The key advantage of \texttt{Falcon} is achieving a larger reduction while satisfying the quality requirement with high probability.}
To evaluate this advantage, we compare all baselines from:

\begin{itemize}[left=0pt]
    \item \emph{Quality Satisfaction} -- How closely and reliably does each method meet the target quality? We report the average F1 achieved under each target \(\alpha\), ranging from 0.80 to 0.95 in steps of 0.01.
    We also report the quality satisfaction rate, defined as the fraction of evaluated query–target settings whose achieved F1 is at least \(\alpha\).
    \item \emph{Oracle Reduction} -- How many oracle LLM calls does each method save? We report the reduction rate as the fraction of documents labeled without the oracle. Sampled documents are labeled by the oracle and thus do not contribute to the reduction.
    
    \item \emph{Time and Token Cost} -- 
    Time cost is the end-to-end time from a query 
    to labeling the full collection. Token cost includes both the cheap proxy and the oracle. For methods with an offline stage (\aurora, \scaledoc, \texttt{UniCSV}, and \texttt{SimCSV}), we exclude offline embedding time and tokens and discuss them separately.
    \item \emph{Robustness} -- Does a method maintain its quality and reduction, 
    especially in extreme-selectivity regimes? We report the quality satisfaction rate and the reduction rate.
    \item \emph{Proxy Quality} -- How well does the proxy separate positive from negative documents? We report AUC-ROC~\cite{fawcett2006introduction}.
    \item \emph{Ablation} -- What does each component contribute to \aurora? We disable or replace individual designs and measure their effect.

\end{itemize}

\myparagraph{Implementations}
All methods use {Qwen text-embedding-v4}~\cite{qwen3embedding} for document embedding. For \lotus and \bargain, we use {Qwen3.5-0.8B}~\cite{qwen35_08b} as the smaller LLM proxy. All final oracle-labeling tasks are performed by {Qwen3.5-397B}~\cite{qwen35_397b}. 
{For \csv, we set its lower bound \(lb=1-\alpha\) to simulate different quality constraints.} Unless specified, we set the target quality to \(\alpha=0.9\) and the failure probability to \(\delta=0.05\) by default. {Experiments with varying \(\alpha\) use 5 random seeds, while those with a fixed \(\alpha=0.9\) use 100 random seeds.}

\subsection{Overall Quality and Reduction} \label{sec:exp-reduction-quality}
\myparagraph{Achieved F1 and Quality Satisfaction}
Figures~\ref{fig:f1} and~\ref{fig:satisfaction} report the average achieved F1 under varying $\alpha$ and the per-query quality satisfaction rate when $\alpha=0.9$, respectively. We make three observations: 
(1) At the dataset level, \aurora meets all target F1 settings (Figure~\ref{fig:f1}), while avoiding severe per-query violations observed in existing methods, e.g., \lotus, \scaledoc, and \csv on \bank (Figure~\ref{fig:satisfaction}). This contrast shows that meeting the target in average F1 does not necessarily ensure reliable quality satisfaction across queries. Among the 120 queries, only 3 evaluated with \aurora achieve satisfaction rates below 95\%, consistent with its asymptotic quality guarantee, which allows minor violations in finite-sample settings.
(2) Although \texttt{UniCSV} and \texttt{SimCSV}'s achieved F1 generally increases with \(\alpha\), their satisfaction rates remain low on most datasets (Figure~\ref{fig:satisfaction}). 
They control the quality of individual clusters rather than directly certifying the end-to-end F1. Increasing \(\alpha\) 
routes more uncertain clusters to the oracle, but does not ensure that the final collection-level F1 reaches the target. 
(3) \bargain provides finite-sample theoretical guarantees and achieves a 100\% quality satisfaction rate across all queries. However, its lower reduction (Figure~\ref{fig:pure_reduction}) reflects a more conservative quality–reduction trade-off, partly due to its separate precision and recall constraints.

\myparagraph{Reduction Rate}
Figure~\ref{fig:pure_reduction} reports the average reduction rate over queries for which each method satisfies the target F1. We exclude constraint-violating queries because oracle savings obtained by sacrificing output quality do not constitute valid reductions.
We make three observations: 
(1) \aurora achieves the highest reduction on 4 of 6 datasets and remains competitive on the other 2. As the target F1 increases, reduction generally decreases because stricter quality requirements leave fewer documents eligible for automatic labeling. 
(2) \scaledoc obtains relatively modest reductions in our experiments, consistent with the limited score separation and discretized threshold search of its learned-proxy cascade.
(3) Although \csv on \law and \lotus on \mimic achieve higher reductions through cluster-level label propagation and proxy-based routing, respectively, they satisfy the quality constraint on fewer queries and may not offer a better overall trade-off. \aurora provides a more reliable and computationally efficient alternative, as supported by the latency and token results in Tables 3 and 4.



\begin{table*}[t] \small
    \centering
    \caption{Average query-level end-to-end latency (s) at target F1 (\(\alpha=0.90\)).}
    \vspace{-1.25em}
    \label{tab:latency}
    \definecolor{TotalGray}{RGB}{226,226,226}
    \newcommand{\latencyentry}[3]{\makebox[2.2em][c]{(#1)}\makebox[7em][c]{#2\,\(\pm\)\,#3}}
    \newcommand{\offlineentry}[1]{\makebox[9.2em][c]{#1}}
    \renewcommand{\arraystretch}{0.88}
    \setlength{\tabcolsep}{1.4pt}
    \begin{tabular*}{\textwidth}{@{\extracolsep{\fill}}c cccccc}
    \toprule
    \textbf{Methods} & \bank & \emotion & \patent & \law & \med & \mimic \\
    \midrule
    Offline
    & \offlineentry{3.11E+3} & \offlineentry{3.30E+3} & \offlineentry{3.30E+3} & \offlineentry{8.52E+2} & \offlineentry{3.26E+3} & \offlineentry{4.01E+3} \\
    \lotus
    & \latencyentry{9}{2.27E+4}{2.39E+3} & \latencyentry{8}{3.57E+4}{6.44E+3} & \latencyentry{4}{2.81E+4}{3.22E+3} & \latencyentry{2}{9.47E+3}{2.93E+2} & \latencyentry{14}{2.44E+4}{4.28E+3} & \latencyentry{12}{4.16E+4}{1.10E+4} \\
    \bargain
    & \latencyentry{20}{2.58E+4}{1.76E+3} & \latencyentry{20}{3.88E+4}{3.39E+3} & \latencyentry{20}{3.22E+4}{9.76E+3} & \latencyentry{20}{9.79E+3}{5.68E+2} & \latencyentry{20}{2.47E+4}{5.08E+3} & \latencyentry{20}{5.14E+4}{1.19E+4} \\
    \scaledoc
    & \latencyentry{7}{1.24E+4}{6.64E+1} & \latencyentry{11}{2.81E+4}{3.81E+2} & \latencyentry{16}{2.67E+4}{5.96E+3} & \latencyentry{3}{7.35E+3}{5.26E+2} & \latencyentry{18}{2.22E+4}{5.26E+2} & \latencyentry{20}{5.50E+4}{5.52E+2} \\
    \texttt{UniCSV}
    & \latencyentry{4}{9.41E+3}{2.55E+3} & \latencyentry{4}{2.66E+4}{1.32E+2} & \latencyentry{11}{1.24E+4}{9.34E+3} & \latencyentry{5}{2.72E+3}{3.79E+2} & \latencyentry{10}{1.61E+4}{5.40E+3} & \latencyentry{13}{4.83E+4}{6.57E+3} \\
    \texttt{SimCSV}
    & \latencyentry{5}{8.88E+3}{2.53E+3} & \latencyentry{4}{2.67E+4}{1.14E+2} & \latencyentry{11}{1.27E+4}{9.33E+3} & \cellcolor{TotalGray}{\latencyentry{7}{2.62E+3}{3.68E+2}} & \latencyentry{11}{1.61E+4}{5.10E+3} & \latencyentry{13}{4.87E+4}{6.49E+3} \\
    \aurora
    & \cellcolor{TotalGray}{\latencyentry{20}{2.87E+3}{9.37E+2}} & \cellcolor{TotalGray}{\latencyentry{18}{1.21E+4}{3.23E+3}} & \cellcolor{TotalGray}{\latencyentry{20}{8.44E+3}{4.52E+3}} & \latencyentry{19}{4.16E+3}{8.05E+2} & \cellcolor{TotalGray}{\latencyentry{20}{5.56E+3}{2.71E+3}} & \cellcolor{TotalGray}{\latencyentry{20}{2.32E+4}{7.62E+3}} \\
    \bottomrule
    \end{tabular*}
    \vspace{-0.75em}
\end{table*}

\begin{table*}[t] \small
    \centering
    \caption{Average query-level end-to-end token consumption at target F1 (\(\alpha=0.90\)).}
    \vspace{-1.25em}
    \label{tab:token}
    \definecolor{TotalGray}{RGB}{226,226,226}
    \newcommand{\tokenentry}[3]{\makebox[2.2em][c]{(#1)}\makebox[7em][c]{#2\,\(\pm\)\,#3}}
    \newcommand{\offlinetokens}[1]{\makebox[9.2em][c]{#1}}
    \renewcommand{\arraystretch}{0.88}
    \setlength{\tabcolsep}{1.4pt}
    \begin{tabular*}{\textwidth}{@{\extracolsep{\fill}}c cccccc}
    \toprule
    \textbf{Methods} & \bank & \emotion & \patent & \law & \med & \mimic \\
    \midrule
    Offline
    & \offlinetokens{1.48E+5} & \offlinetokens{1.71E+5} & \offlinetokens{1.30E+6} & \offlinetokens{2.24E+6} & \offlinetokens{4.18E+6} & \offlinetokens{1.41E+7} \\
    \lotus
    & \tokenentry{9}{1.21E+6}{1.61E+5} & \tokenentry{8}{1.35E+6}{1.52E+5} & \tokenentry{4}{2.76E+6}{2.35E+5} & \tokenentry{2}{3.59E+6}{1.46E+5} & \tokenentry{14}{6.54E+6}{9.78E+5} & \tokenentry{12}{2.14E+7}{3.09E+6} \\
    \bargain
    & \tokenentry{20}{1.45E+6}{1.31E+5} & \tokenentry{20}{1.47E+6}{1.19E+5} & \tokenentry{20}{3.05E+6}{6.80E+5} & \tokenentry{20}{4.44E+6}{1.54E+6} & \tokenentry{20}{6.58E+6}{1.18E+6} & \tokenentry{20}{2.37E+7}{3.33E+6} \\
    \scaledoc
    & \tokenentry{7}{8.23E+5}{2.70E+4} & \tokenentry{11}{8.28E+5}{4.52E+4} & \tokenentry{16}{1.85E+6}{4.20E+5} & \tokenentry{3}{1.81E+6}{2.28E+5} & \tokenentry{18}{4.88E+6}{1.21E+5} & \tokenentry{20}{1.52E+7}{1.06E+5} \\
    \texttt{UniCSV}
    & \tokenentry{4}{6.43E+5}{1.57E+5} & \tokenentry{4}{8.29E+5}{6.81E+4} & \tokenentry{11}{8.68E+5}{6.58E+5} & \tokenentry{5}{8.90E+5}{3.03E+5} & \tokenentry{10}{3.61E+6}{1.24E+6} & \tokenentry{13}{1.34E+7}{1.68E+6} \\
    \texttt{SimCSV}
    & \tokenentry{5}{6.11E+5}{1.55E+5} & \tokenentry{4}{8.31E+5}{6.76E+4} & \tokenentry{11}{8.87E+5}{6.57E+5} & \cellcolor{TotalGray}{\tokenentry{7}{8.60E+5}{2.84E+5}} & \tokenentry{11}{3.60E+6}{1.17E+6} & \tokenentry{13}{1.35E+7}{1.65E+6} \\
    \aurora
    & \cellcolor{TotalGray}{\tokenentry{20}{1.99E+5}{6.50E+4}} & \cellcolor{TotalGray}{\tokenentry{18}{3.56E+5}{8.83E+4}} & \cellcolor{TotalGray}{\tokenentry{20}{5.92E+5}{3.20E+5}} & \tokenentry{19}{1.41E+6}{6.00E+5} & \cellcolor{TotalGray}{\tokenentry{20}{1.25E+6}{6.15E+5}} & \cellcolor{TotalGray}{\tokenentry{20}{6.59E+6}{2.22E+6}} \\
    \bottomrule
    \end{tabular*}
    \vspace{-1.50em}
\end{table*}


\subsection{Time and Token Cost}\label{sec:exp-cost}
We divide the workflow into an \emph{offline stage} and an \emph{online stage}. The offline stage generates query-agnostic document embeddings once per dataset, which are reused by embedding-based methods (\aurora, \scaledoc, and \texttt{CSV}). The online stage covers the process from query arrival to assigning labels to all documents.
To avoid rewarding methods that sacrifice quality for reduction, Tables~\ref{tab:latency} and \ref{tab:token} report latency and token costs only for queries achieving at least 95\% quality satisfaction across 100 random seeds. The parenthesized number preceding each entry indicates the number of queries included in the statistics. The best result is highlighted in light gray.


We make four observations: 
(1) \aurora achieves the lowest latency and token cost on 5 of 6 datasets, remaining competitive on \law. Against \lotus and \bargain, it reduces online latency by up to \(7.9\times\) and \(9.0\times\), and token consumption by \(6.1\times\) and \(7.3\times\), respectively. These gains arise from lightweight proxy scoring that avoids per-document small LLM inference and cascade routing that reduces oracle calls.
(2) Although \lotus achieves higher oracle-call reduction on \mimic, it does not achieve the lowest latency or token cost due to per-document inference overhead of its LLM-based proxy.
(3) On \law, only 5 and 7 queries from \texttt{UniCSV} and \texttt{SimCSV} meet the quality requirement, versus 19 for \aurora. Despite their lower latency and token costs on these limited subsets, \aurora achieves broader quality-compliant coverage with competitive costs, offering a better quality–efficiency trade-off.
(4) Offline embedding generation incurs one-time latency and token costs comparable to those of a single online query. Since embeddings are reused across queries on the same dataset, these costs are amortized and become less significant as more queries are processed.

\subsection{Robustness} \label{sec:exp_robustness}
We categorize the 120 queries from six datasets into three groups based on predicate selectivity: \emph{Moderate} (selectivity: 20\%–80\%, 40 queries), \emph{Skewed} (selectivity: 5\%–20\% or 80\%–95\%, 51 queries), and \emph{Extreme} (selectivity: below 5\% or above 95\%, 29 queries). Figure~\ref{fig:robustness}(a) and (b) report the per-query quality satisfaction rate and average reduction rate across these groups, respectively.

\begin{figure}[t]
  \centering
  \includegraphics[width=0.48\textwidth]{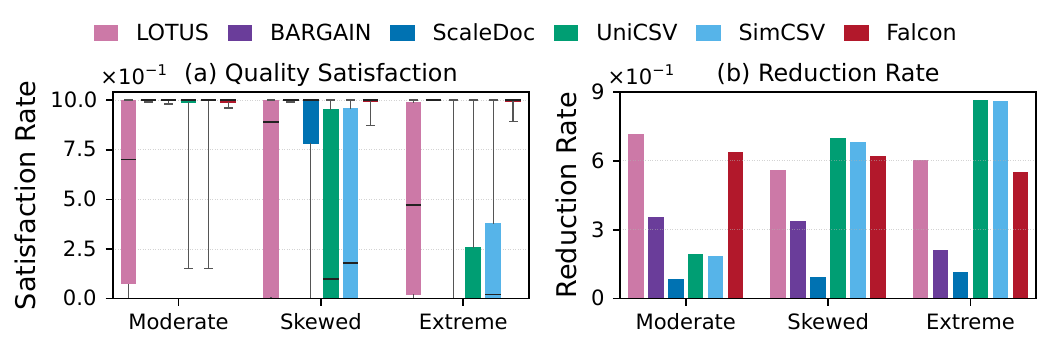}
  \vspace{-2.75em}
  \caption{Per-query quality satisfaction and oracle-call reduction across different predicate selectivities ($\alpha=0.9$).}
  \vspace{-0.75em}
  \label{fig:robustness}
\end{figure}

We make two observations: 
(1) \aurora maintains consistently high quality satisfaction across all selectivity regimes, with only a modest reduction decline from moderate to extreme selectivity. This robustness benefits from its selectivity-aware sampling and iterative proxy refinement, which improve minority-class coverage and score discrimination.
(2) Baselines exhibit less balanced quality–reduction trade-offs. \bargain maintains high satisfaction but achieves lower reduction, while \lotus and \scaledoc show less stable quality satisfaction. Although \texttt{UniCSV} and \texttt{SimCSV} achieve higher reductions under skewed and extreme selectivity, these results are limited to quality-satisfying queries, whereas their overall satisfaction rates remain low. In contrast, \aurora offers a more consistent trade-off across selectivity regimes.

\subsection{Proxy Quality}\label{sec:proxy quality}
Figure~\ref{fig:auc} demonstrates the AUC-ROC of the proxy scores of \lotus, \scaledoc, and \aurora (1 indicates perfect separation).  We omit \bargain and \texttt{CSV} because they do not produce document-level proxy scores.  
We make two observations:
(1) \aurora achieves the highest median AUC-ROC on five of the six datasets and exhibits consistently small variation across predicates. This shows that its proxy ranks positive documents above negative ones reliably across different domains. 
(2) \lotus degrades substantially on \law. Legal documents often share similar terminology and case context, regardless of whether they satisfy the predicate. Because the small-LLM proxy in \lotus is not adapted using predicate-specific oracle labels, it often assigns high scores to topically related but predicate-negative documents. In contrast, \scaledoc and \aurora adapt their scores using labeled examples, enabling them to suppress these false positives and maintain higher AUC-ROC. 

\begin{figure}[t]
  \centering
  \includegraphics[width=0.48\textwidth]{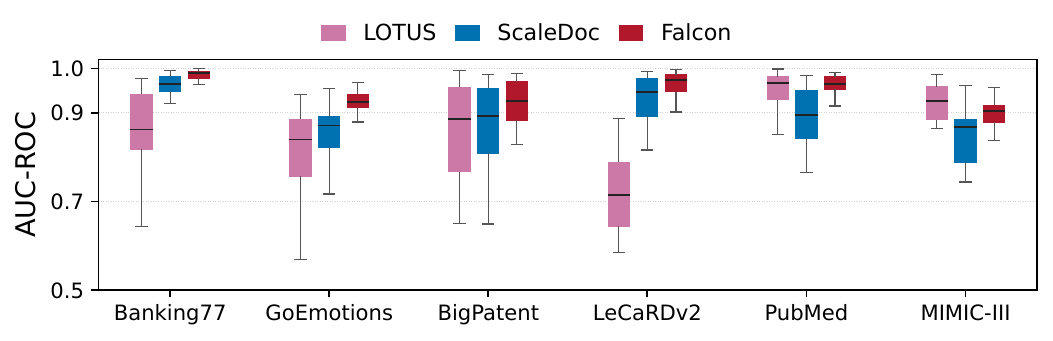}
  \vspace{-2.5em}
  \caption{Proxy-score AUC-ROC across datasets.}
  \label{fig:auc}
\end{figure}

\subsection{Ablation Studies}
\myparagraph{Contribution of the Overall Design}
We construct three \aurora variants to evaluate the designs in Sections~\ref{sec:pred-vec-gen-main}--\ref{sec:th-select-wg}. \aurora-\texttt{Doc} directly uses the pretrained document embeddings, removing the predicate-aware semantic vector construction in Section~\ref{sec:pred-vec-gen-main}. \aurora-\texttt{Logistic} replaces the multi-view scoring and residual correction with a standard logistic regression model. \aurora-\texttt{UniBgt} replaces the certified threshold-selection method in Section~\ref{sec:th-select-wg} with uniform audit sampling, and selects a threshold satisfying \(\widehat{\mathcal{B}}_\alpha(\tau) \geq 0\) without lower confidence bound certification.
Figure~\ref{fig:abl_overall} compares their quality–reduction trade-offs and execution costs. We observe that:
(1) Overall, \aurora achieves the best overall balance across quality satisfaction, oracle-call reduction, latency, and token consumption, showing that the three major design components contribute complementary benefits. 
(2) \aurora-\texttt{UniBgt} selects overly aggressive thresholds because it relies only on estimated F1 budgets without accounting for sampling uncertainty. Although it achieves the highest reduction and the lowest latency and token consumption, its low quality satisfaction rate makes these efficiency gains invalid.

\begin{figure}[t]
  \centering
  \includegraphics[width=0.48\textwidth]{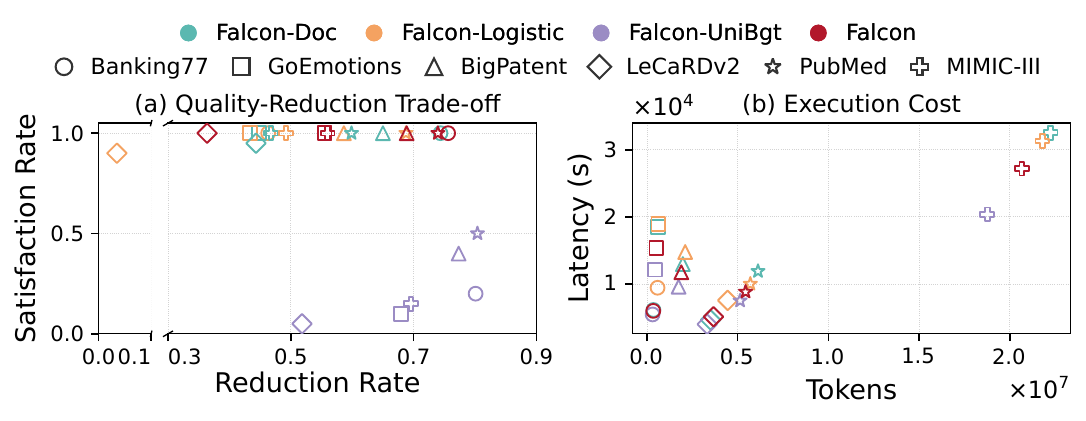}
  \vspace{-2.5em}
  \caption{Ablation of \aurora's major components ($\alpha=0.9$).}
  \vspace{-1.5em}
  \label{fig:abl_overall}
\end{figure}

\begin{figure}[t]
  \centering
  \includegraphics[width=0.48\textwidth]{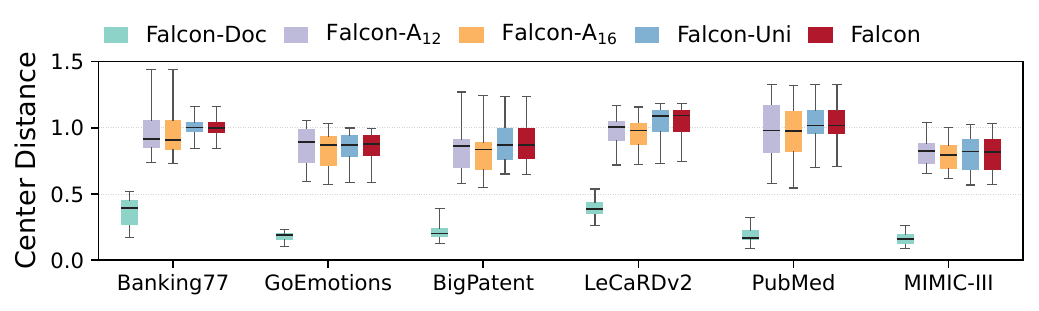}
  \vspace{-2.5em}
  \caption{Centroid distance under different vector variants.}
  \vspace{-0.75em}
  \label{fig:abl_sec_3}
\end{figure}

\myparagraph{
Predicate-aware Semantic Vector}
Figure~\ref{fig:abl_sec_3} compares how well semantic vectors constructed by different variants separate positive and negative documents. Using oracle labels, we measure this separation by the distance between positive and negative centroids in the resulting vector space. The evaluated methods include: \aurora-\texttt{Doc}, which directly uses the pretrained document embeddings; \aurora-\texttt{A}\(_{12}\) and \aurora-\texttt{A}\(_{16}\), which retain the top 12 and 16 candidate directions, respectively (Section~\ref{sec:di-select};  \aurora-\texttt{Uni}), which replaces tail-guided sampling with uniform sampling. 

We observe:
(1) All predicate-aware variants achieve larger centroid distances than \aurora-\texttt{Doc}, showing that predicate-specific directions substantially improve class separation. However, retaining more directions does not always help. \aurora-\texttt{A}\(_{16}\) can produce weaker separation because lower-ranked directions contain less discriminative information and may introduce noise.
(2) \aurora-\texttt{Uni} and \aurora achieve larger or more stable center distances, demonstrating the effectiveness of the Fisher direction design introduced in Section~\ref{sec:di-select}. In particular, \aurora achieves a high median center distance across all datasets, indicating that the samples acquired through tail-guided sampling facilitate the construction of a highly discriminative vector space.
(3) Compared with \aurora-\texttt{Uni}, \aurora achieves higher mean centroid distances and more stable results across all datasets. This improvement stems from tail-guided sampling, which acquires more minority-class samples and thereby enables more reliable construction of predicate directions, ultimately yielding a more discriminative semantic vector space.

\begin{figure}[t]
  \centering
  \includegraphics[width=0.48\textwidth]{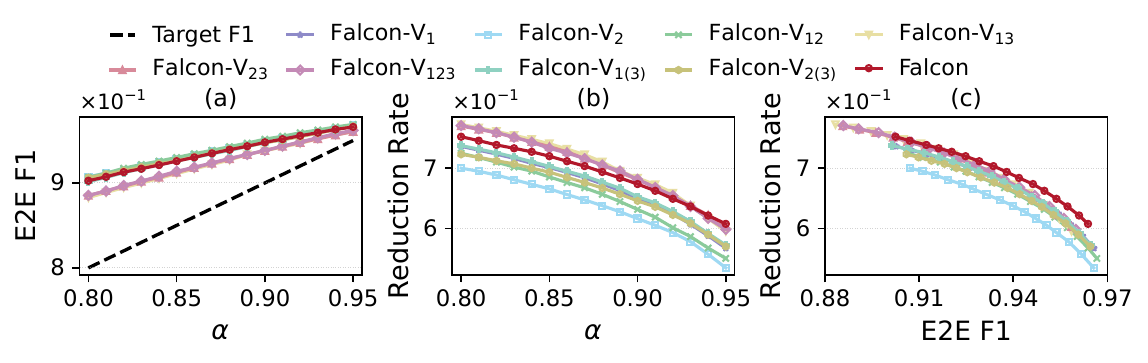}
  \vspace{-2em}
  \caption{Impact of views on F1 and oracle-call reduction.}
  \vspace{-1.0em}
  \label{fig:view}
\end{figure}

\begin{figure}[t]
  \centering
  \includegraphics[width=0.48\textwidth]{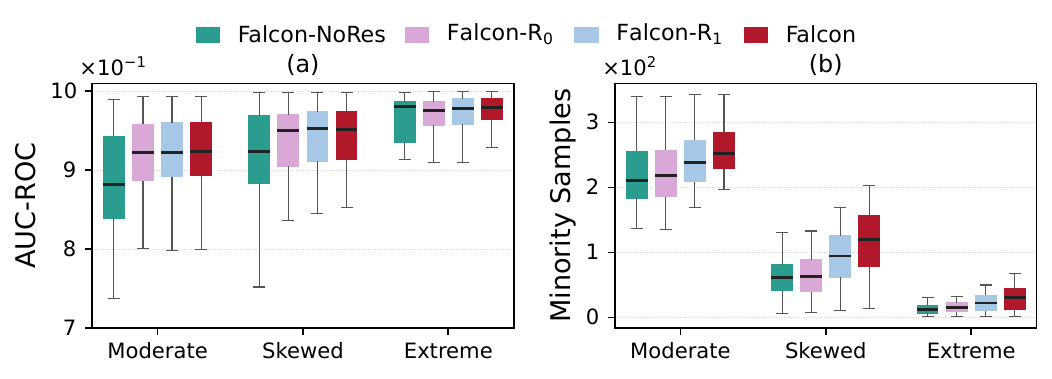}
  \vspace{-2em}
  \caption{Impact of residual correction and label refresh.}
  \vspace{-1.0em}
  \label{fig:refresh}
\end{figure}

\myparagraph{
Proxy Construction} 
We first evaluate \aurora under different view combinations in Figure~\ref{fig:view} (using 5 random seeds for each $\alpha$). View $1$, View $2$, and View $3$ denote the direction, neighborhood, and semantic-prior views, respectively. The suffix of each variant indicates the views it uses. A view enclosed in parentheses is activated only as a fallback; for example, \aurora-V\(_{12(3)}\) combines the direction and neighborhood views and selectively activates the semantic-prior view.
No single view performs best in both achieved F1 and oracle-call reduction. A view may improve one metric while degrading the other. Combining complementary views produces a better quality–reduction trade-off, with \aurora remaining close to the target F1 while achieving high reduction and lying on the Pareto frontier in Figure~\ref{fig:view}(c).

We further evaluate residual correction and iterative refreshment in Figure~\ref{fig:refresh} ($\alpha=0.9$), using the same query grouping as in Section~\ref{sec:exp_robustness}. \aurora-\texttt{NoRes} removes residual correction. \aurora-\texttt{R}$_0$ directly samples \(7\%\) documents in tail-guided sampling without refreshment. \aurora-\texttt{R}$_1$ samples \(5\%\) documents in tail-guided sampling and adds  \(2\%\) during one refreshment round.
\aurora consistently achieves the highest AUC-ROC and acquires more minority-class labels across the selectivity regimes. 
Without the correction, the less discriminative scores weaken guidance for label refresh, reducing the chance of acquiring minority-class documents. Additional refresh rounds further improve minority-class coverage, 
particularly under skewed and extreme predicate selectivity.

\begin{figure}[t]
  \centering
  \includegraphics[width=0.48\textwidth]{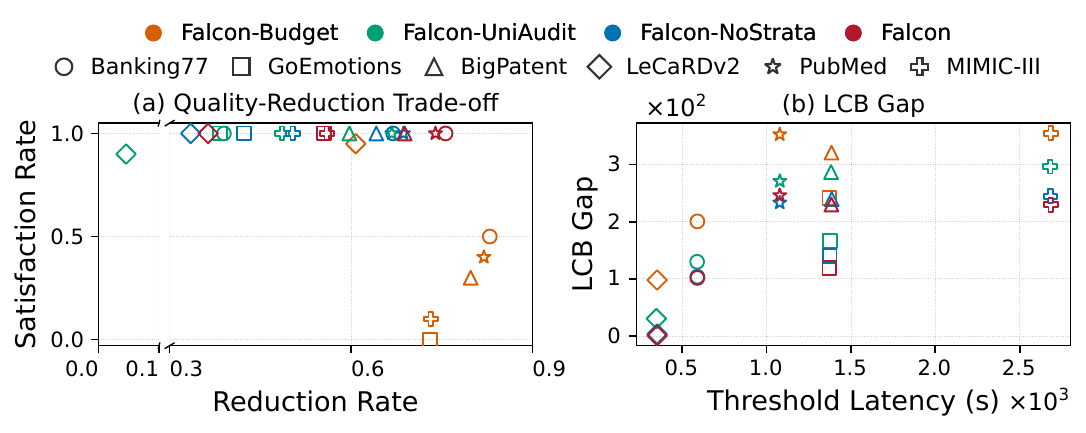}
  \vspace{-2em}
  \caption{Ablation of \aurora's certified threshold selection.}
  \vspace{-0.5em}
  \label{fig:abl_section5}
\end{figure}

\myparagraph{
Threshold Selection} 
Figure~\ref{fig:abl_section5} evaluates three variants of \aurora ($\alpha=0.9$). \aurora-\texttt{Budget} selects the threshold pair with the largest reduction subject to \(\widehat{\mathcal{B}}_\alpha(\tau) \geq 0\). \aurora-\texttt{UniAudit} selects audit samples through uniform sampling. \aurora-\texttt{NoStrata} constructs the budget estimate and confidence bound using only the score bins, without the frontier-aware strata. 
The LCB gap denotes the difference between the empirical F1 budget and its lower confidence bound at the selected threshold \(\tau\).
We make two observations: 
(1) \aurora-\texttt{Budget} and \aurora-\texttt{NoStrata} exhibit lower constraint satisfaction rates. This is because \aurora-\texttt{Budget} ignores sampling uncertainty and therefore tends to select overly aggressive thresholds. \aurora-\texttt{NoStrata} fails to capture differences in document behavior across candidate thresholds. It can underestimate the uncertainty of the budget estimate.
(2) \aurora achieves a better quality–reduction trade-off than \aurora-\texttt{UniAudit}. Its variance-aware audit sampling allocates more labels to score ranges whose contributions are harder to estimate, allowing thresholds closer to the feasibility boundary to be safely selected.

\begin{figure}[t]
  \centering
  \includegraphics[width=0.48\textwidth]{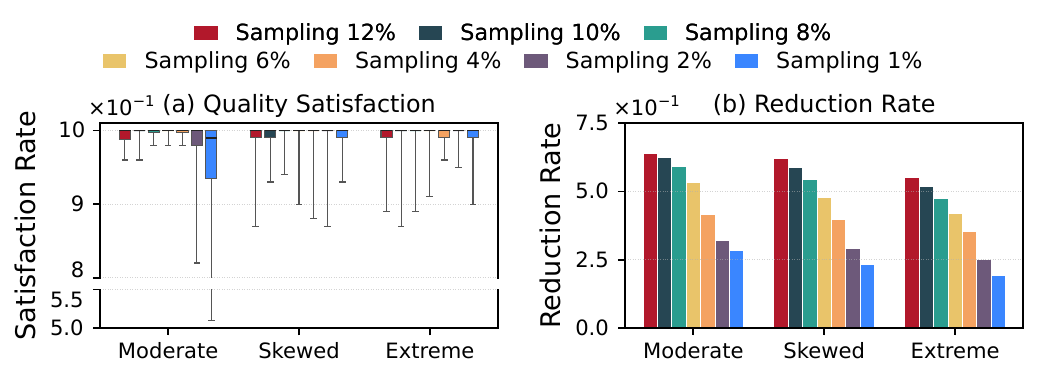}
  \vspace{-2.5em}
  \caption{\aurora's satisfaction rate and reduction rate across predicate selectivities under varying sampling budgets.}
  \vspace{-0.5em}
  \label{fig:exp_sampling}
\end{figure}

\subsection{Impact of Sampling Budget}
Figure~\ref{fig:exp_sampling} reports the quality satisfaction rate and oracle-call reduction under varying sampling budgets, with target F1 ($\alpha$=0.9) and 100 random seeds per query. We use the same selectivity-based query grouping as in Section~\ref{sec:exp_robustness}. By default, \aurora allocates 12\% of documents to oracle sampling; smaller budgets proportionally reduce sampling across all stages.
For example, under a 6\% sampling budget, the number of samples used in each stage is halved relative to the default configuration.
We make three observations:
(1) \aurora maintains high quality satisfaction across selectivity regimes and sampling budgets, with median satisfaction rates close to 100\% (Figure~\ref{fig:exp_sampling}(a)). However, smaller budgets increase variability across queries, particularly at 1\%, where insufficient labels can weaken proxy refinement and quality estimation. Notably, the medians remain high even at the smallest budgets, indicating that quality degradation is not uniform across queries. Instead, the extended lower whiskers suggest sensitivity concentrated in a few difficult queries, highlighting the importance of adequate rare-class and audit coverage.
(2) As the sampling budget decreases from 12\% to 1\%, reduction consistently declines across all selectivity regimes (Figure~\ref{fig:exp_sampling}(b)). Although smaller budgets reduce sampling overhead, the resulting loss of supervision and increased uncertainty in F1-budget estimation limit the documents that can be reliably resolved without the oracle.
(3) Extreme-selectivity queries generally achieve lower reductions than moderate-selectivity queries, despite maintaining high quality satisfaction. Rare-class scarcity limits informative supervision and makes aggressive routing harder to justify under the F1 constraint. Consequently, \aurora tends to retain more documents for oracle evaluation, prioritizing quality over additional reduction.
\section{Related Work}


\myparagraph{LLM-Powered Data Systems}
We categorize LLM-powered data systems into two groups according to the input format used for task specification:
(1) One line of work exposes structured or semi-structured query interfaces, where users specify semantic computation as SQL-like queries, operator pipelines, or semantic-operator directed acyclic graphs (DAGs). These systems focus on optimizing the semantic operator-based plans. For example, \lotus \cite{Patel25lotus}, \texttt{Palimpzest} \cite{liu2025palimpzest}, \texttt{DocETL} \cite{shankar2025docetl}, and \texttt{Sema} \cite{qi2026sema} provide declarative semantic operators and runtime optimization, while \texttt{Nirvana} \cite{zhu2025nirvana} and \texttt{Abacus} \cite{russo2026abacus} further study semantic-aware logical rewriting and cost-based physical optimization over semantic-operator plans. 
(2) Another line of work accepts natural-language semantic inputs and translates them into executable data-access or reasoning procedures. \texttt{Pneuma} \cite{balaka2025pneuma} supports tabular data discovery with LLM-based representations; \texttt{ST-Raptor} \cite{tang2025straptor} studies natural-language question answering over semi-structured tables, while \texttt{ReasonTabQA} \cite{pan2026reasontabqa} benchmarks complex table reasoning in industrial scenarios; and \texttt{AgenticScholar} \cite{lan2026agenticscholar} translates natural-language scholarly queries into executable DAG plans over scholarly corpora. Beyond execution optimization, these systems also study how to translate and organize natural-language requests into correct executable plans.

\myparagraph{Cost-efficient Semantic Filtering}
A central challenge in semantic operators is cost: naive execution invokes the oracle LLM on every record, making semantic filtering and joining expensive at large scale. Prior works reduce this cost through proxy-based cascades, sampling, and clustering-based label propagation. For semantic filtering, \lotus \cite{Patel25lotus} and \bargain \cite{Zeighami25bargain} instantiate this idea with cheap-LLM proxies: they use affordable LLMs to score documents and adopt SUPG-style \cite{kang2020supg} thresholding to directly assign labels to high-confidence documents. 
Other studies show that embedding-based methods can substantially reduce the cost and latency of semantic filtering operators \cite{chung2026aiquery,zhang2026scaledoc,hou26csv,kim2026fast}. For example, \scaledoc \cite{zhang2026scaledoc} learns an online proxy with contrastive learning to score whether each document satisfies the predicate; \csv \cite{hou26csv} breaks the linear invocation pattern through clustering, intra-cluster sampling, and label propagation; and recent work further combines and unifies clustering-based and proxy-learning approaches as different design points of fast semantic filtering \cite{kim2026fast}.
Semantic joins face an analogous cost challenge because pairwise LLM comparisons can grow quadratically with the input sizes. Existing systems \cite{Patel25lotus,russo2026abacus,zhu2025nirvana} reduce this cost through candidate generation, semantic-aware plan optimization, and cost-based physical implementation selection.

\vspace{-0.25em}
\section{Conclusion}
In this paper, we study the problem of oracle-efficient semantic filtering over document collections. We introduce \aurora, a fast, lightweight, and robust proxy cascade that transforms generic document embeddings into predicate-aware semantic vectors, progressively refines a multi-view proxy using a limited oracle-label budget, and certifies cascade thresholds through an additive F1 budget. \aurora further uses guided label acquisition to remain effective under extreme predicate selectivity. Our extensive experiments show that \aurora achieves up to \(82\%\) oracle-call reduction and reduces end-to-end latency and token consumption by up to $7.3\times$, while closely meeting the target F1 requirement.


\newpage


\bibliographystyle{ACM-Reference-Format}
\bibliography{sample}

\newpage
\appendix
\section{Derivation of the Additive F1 Budget}
\label{app:f1-budget}

We show that the F1 constraint under a threshold pair can be expressed as a sum of per-document budget contributions.

Consider a threshold pair $\tau=(l,r)$ and a multi-view score
$\ell_i^{\mathrm{mv}}(\mathbf{w})$. Define:
\begin{align}
L_i(\tau;\mathbf{w})
&=
\mathbb{I}\!\left[
\ell_i^{\mathrm{mv}}(\mathbf{w}) \leq l
\right], \\
H_i(\tau;\mathbf{w})
&=
\mathbb{I}\!\left[
\ell_i^{\mathrm{mv}}(\mathbf{w}) \geq r
\right].
\end{align}
A document with $L_i=1$ is resolved as negative, one with $H_i=1$
is resolved as positive, and one with $L_i=H_i=0$ is routed to the
oracle and receives its oracle label. Since $l<r$, $L_i$ and $H_i$
cannot both equal one. The final predicted label can therefore be
written as
\begin{equation}
\hat{y}_i
=
H_i(\tau;\mathbf{w})
+
\left(
1-L_i(\tau;\mathbf{w})-H_i(\tau;\mathbf{w})
\right)y_i.
\end{equation}

For brevity, we omit $(\tau;\mathbf{w})$ from $L_i$ and $H_i$
below. The numbers of true positives, false negatives, and false
positives produced by the cascade are
\begin{align}
\mathrm{TP}
&=
\sum_{i=1}^{n} y_i\hat{y}_i
=
\sum_{i=1}^{n} y_i(1-L_i), \\
\mathrm{FN}
&=
\sum_{i=1}^{n} y_i(1-\hat{y}_i)
=
\sum_{i=1}^{n} L_i y_i, \\
\mathrm{FP}
&=
\sum_{i=1}^{n} (1-y_i)\hat{y}_i
=
\sum_{i=1}^{n} H_i(1-y_i).
\end{align}
A positive document is predicted as negative only when it falls
below the lower threshold. Similarly, a negative document becomes
a false positive only when it exceeds the upper threshold.
Documents routed to the oracle introduce no classification error.

For a target quality level $\alpha\in(0,1]$, the F1 requirement is
\begin{equation}
\mathrm{F1}
=
\frac{2\mathrm{TP}}
{2\mathrm{TP}+\mathrm{FP}+\mathrm{FN}}
\geq \alpha.
\end{equation}
Assuming that the denominator is positive, multiplying both sides
by the denominator gives
\begin{equation}
2\mathrm{TP}
\geq
\alpha
\left(
2\mathrm{TP}+\mathrm{FP}+\mathrm{FN}
\right),
\end{equation}
which is equivalent to
\begin{equation}
2(1-\alpha)\mathrm{TP}
-
\alpha\mathrm{FP}
-
\alpha\mathrm{FN}
\geq 0.
\end{equation}

Substituting the expressions for $\mathrm{TP}$, $\mathrm{FP}$, and
$\mathrm{FN}$ yields
\begin{align}
&\quad
2(1-\alpha)
\sum_{i=1}^{n} y_i(1-L_i)
-
\alpha
\sum_{i=1}^{n} H_i(1-y_i)
-
\alpha
\sum_{i=1}^{n} L_i y_i
\nonumber\\
&=
\sum_{i=1}^{n}
\left[
2(1-\alpha)y_i
-
(2-\alpha)L_i y_i
-
\alpha H_i(1-y_i)
\right]
\geq 0.
\end{align}

We therefore define the per-document F1-budget contribution as
\begin{equation}
Z_i(\alpha,\tau;\mathbf{w})
=
2(1-\alpha)y_i
-
(2-\alpha)L_i(\tau;\mathbf{w})y_i
-
\alpha H_i(\tau;\mathbf{w})(1-y_i),
\end{equation}
and the collection-level F1 budget as
\begin{equation}
B_\alpha(\tau;\mathbf{w})
=
\sum_{i=1}^{n}
Z_i(\alpha,\tau;\mathbf{w}).
\end{equation}
Consequently,
\begin{equation}
\mathrm{F1}\geq\alpha
\quad\Longleftrightarrow\quad
B_\alpha(\tau;\mathbf{w})\geq 0.
\end{equation}

This additive form makes each document’s contribution to the transformed F1 constraint explicit. It
allows the collection-level F1 requirement to be estimated from
sampled oracle labels.
\section{Unbiasedness of the Pilot Budget Estimator}
\label{app:budget-estimator}

This appendix proves the unbiasedness of the provisional F1-budget
estimator used to construct the candidate frontier in
Section~\ref{sec:PF-gen}. The proxy-construction sample $S$ is fully
labeled. The remaining documents $D\setminus S$ are partitioned into
the nonempty score bins ${B}^{\mathrm{score}}$, and the pilot sample
from each bin is used to estimate the contribution of that entire bin.

Fix a threshold pair $\tau$ and condition on the corrected scores and
the finite collection of oracle labels. Thus, for this proof,
$\mathcal{Z}_i(\alpha,\tau)$ is a fixed finite-population quantity.
The provisional estimator in Eq.~\eqref{eq:frontier-budget-signal} is:
\begin{equation}
\widehat{\mathcal{B}}_{\alpha}(\tau)
  =
  \sum_{i\in S}\mathcal{Z}_i(\alpha,\tau)
  +
  \sum_{b\in {B}^{\mathrm{score}}}
  \frac{|b|}{m_b^{(0)}}
  \sum_{i\in S_{a,b}^{(0)}}\mathcal{Z}_i(\alpha,\tau),
\label{eq:appendix-pilot-estimator}
\end{equation}
where $S_{a,b}^{(0)}$ is a uniform sample without replacement of
$m_b^{(0)}=\lceil 1\%\cdot |b|\rceil$ documents from $b$.

\begin{lemma}[Unbiased pilot budget estimate]
For every fixed threshold pair $\tau$,
\begin{equation}
\mathbb{E}\left[
  \widehat{\mathcal{B}}_{\alpha}(\tau)
\right]
  =
  \mathcal{B}_{\alpha}(\tau).
\label{eq:appendix-pilot-unbiased}
\end{equation}
\end{lemma}

\begin{proof}
Consider one nonempty score bin $b$. Since the pilot sample is a
simple random sample without replacement, every document $i\in b$ has
the same inclusion probability
\begin{equation}
\Pr\left(i\in S_{a,b}^{(0)}\right)
  =
  \frac{m_b^{(0)}}{|b|}.
\label{eq:appendix-pilot-inclusion}
\end{equation}
Therefore, by linearity of expectation,
\begin{align}
&\mathbb{E}\left[
  \frac{|b|}{m_b^{(0)}}
  \sum_{i\in S_{a,b}^{(0)}}
  \mathcal{Z}_i(\alpha,\tau)
\right]
\\
&\quad =
  \frac{|b|}{m_b^{(0)}}
  \sum_{i\in b}
  \mathcal{Z}_i(\alpha,\tau)
  \Pr\left(i\in S_{a,b}^{(0)}\right)
\\
&\quad =
  \sum_{i\in b}\mathcal{Z}_i(\alpha,\tau).
\label{eq:appendix-bin-unbiased}
\end{align}

The first term in Eq.~\eqref{eq:appendix-pilot-estimator} contains the
contributions of all documents in $S$, so it is observed without
sampling error. Summing Eq.~\eqref{eq:appendix-bin-unbiased} over the
score bins gives
\begin{align}
\mathbb{E}\left[
  \widehat{\mathcal{B}}_{\alpha}(\tau)
\right]
&=
  \sum_{i\in S}\mathcal{Z}_i(\alpha,\tau)
  +
  \sum_{b\in {B}^{\mathrm{score}}}
  \sum_{i\in b}\mathcal{Z}_i(\alpha,\tau) \\
&=
  \sum_{i\in D}\mathcal{Z}_i(\alpha,\tau)
  =
  \mathcal{B}_{\alpha}(\tau),
\end{align}
where the second equality follows because ${B}^{\mathrm{score}}$
partitions $D\setminus S$. This proves the proposition.
\end{proof}

The proposition is a pointwise statement for each fixed candidate
$\tau$. The Pareto frontier is subsequently selected using these
pilot estimates, so the result should not be interpreted as claiming
that an estimate evaluated at a data-dependent selected threshold is
unbiased after conditioning on that selection event. The post-frontier
certification sample and the simultaneous lower confidence bounds in
Section~\ref{sec:th-select} provide the basis for certifying the
realized frontier candidates.

\section{Proof of the Simultaneous Lower Confidence Bound}
\label{app:proof-lcb}

Once the corrected scores, the audit samples, and the candidate frontier $\mathcal{T}$ have been determined, we fix a target quality level $\alpha$ and consider an arbitrary threshold pair $\tau\in\mathcal{T}$. The audit-sampling and frontier-aware stratification procedures described in Section~\ref{sec:th-select-wg} control the overall audit fraction and merge sparse groups to provide sufficient within-group samples. Under the resulting finite-population regularity conditions, we can assume that the stratified finite-population estimator satisfies the central limit approximation:
\begin{equation}
\label{eq:app-lcb-clt}
  \frac{
    \widehat{\mathcal{B}}_\alpha(\tau)-\mathcal{B}_\alpha(\tau)
  }{
    \sqrt{\widehat{\operatorname{Var}}[
      \widehat{\mathcal{B}}_\alpha(\tau)]}
  }
  \Rightarrow
  \mathcal{N}(0,1),
\end{equation}
where $X\sim\mathcal{N}(0,1)$ denotes the standard normal limit.

We next derive the pointwise lower confidence bound, and then prove
Lemma~\ref{lem:simultaneous-lcb} by applying a union bound.

\begin{lemma}[Pointwise lower confidence bound]
\label{lem:app-pointwise-lcb}
For any fixed $\tau\in\mathcal{T}$, the central limit approximation in
Eq.~\eqref{eq:app-lcb-clt} yields the following asymptotic pointwise bound:
\begin{equation}
\label{eq:app-single-candidate-bound}
  \limsup_{|D|\to\infty}\mathbb{P}\!\left[
    \mathcal{B}_\alpha(\tau)
    <
    \operatorname{LCB}_\alpha(\tau)
  \right]
  \leq
  \frac{\delta}{|\mathcal{T}|},
\end{equation}
where $|D|$ denotes the item number in collection $D$.
\end{lemma}

\begin{proof}
Set
\begin{equation}
\label{eq:app-lcb-threshold}
  x_{\mathcal{T},\delta}
  =
  \sqrt{2\log\!\left(\frac{3|\mathcal{T}|}{\delta}\right)}.
\end{equation}
The radius defined in Eq.~\eqref{eq:stratified-radius} can be written as:
\begin{equation}
\label{eq:app-lcb-radius}
  \operatorname{Rad}_\alpha(\tau)
  =
  x_{\mathcal{T},\delta}
  \sqrt{\widehat{\operatorname{Var}}[
    \widehat{\mathcal{B}}_\alpha(\tau)]}.
\end{equation}
To connect the LCB failure event with the central limit approximation, define
the standardized estimation error:
\begin{equation}
\label{eq:app-standardized-error}
  X_D
  =
  \frac{
    \widehat{\mathcal{B}}_\alpha(\tau)-\mathcal{B}_\alpha(\tau)
  }{
    \sqrt{\widehat{\operatorname{Var}}[
      \widehat{\mathcal{B}}_\alpha(\tau)]}
  }.
\end{equation}

Using the LCB definition and Eq.~\eqref{eq:app-lcb-radius}, the failure event
can be rewritten as:
\begin{align*}
  \left\{
    \mathcal{B}_\alpha(\tau)<\operatorname{LCB}_\alpha(\tau)
  \right\}
  &=\notag
  \left\{
    \mathcal{B}_\alpha(\tau)
    <
    \widehat{\mathcal{B}}_\alpha(\tau)
    -\operatorname{Rad}_\alpha(\tau)
  \right\}
  \\
  &=
  \left\{X_D>x_{\mathcal{T},\delta}\right\}.
\end{align*}

The central limit approximation in Eq.~\eqref{eq:app-lcb-clt}, together with
the Portmanteau theorem, gives:
\begin{align}
  \limsup_{|D|\to\infty}\mathbb{P}\!\left[
    \mathcal{B}_\alpha(\tau)<\operatorname{LCB}_\alpha(\tau)
  \right]
  &=
  \limsup_{|D|\to\infty}\mathbb{P}
  [X_D>x_{\mathcal{T},\delta}]\notag
  \\
  &\leq
  \mathbb{P}[X\geq x_{\mathcal{T},\delta}]
  =
  \mathbb{P}[X>x_{\mathcal{T},\delta}],
\label{eq:app-pointwise-limit-bound}
\end{align}
The equality follows from the continuity of the standard normal distribution.
For $X\sim\mathcal{N}(0,1)$, we use the Gaussian tail bound:
\begin{equation}
\label{eq:app-gaussian-tail}
  \mathbb{P}[X>x]
  \leq
  \exp\!\left(-\frac{x^2}{2}\right),
\end{equation}
Here, $\exp(\cdot)$ denotes the exponential function. Substituting
$x=x_{\mathcal{T},\delta}$ into Eq.~\eqref{eq:app-gaussian-tail} gives:
\begin{align}
  \mathbb{P}[X>x_{\mathcal{T},\delta}]
  &\leq
  \exp\!\left(-\frac{x_{\mathcal{T},\delta}^2}{2}\right)
  =
  \frac{\delta}{3|\mathcal{T}|}
  \leq
  \frac{\delta}{|\mathcal{T}|}.
\label{eq:app-pointwise-tail-bound}
\end{align}
\end{proof}

\begin{proof}[Proof of Lemma~\ref{lem:simultaneous-lcb}]
For each $\tau\in\mathcal{T}$, let $E_\tau$ denote the pointwise LCB failure event:
\[
  E_\tau
  =
  \left\{
    \mathcal{B}_\alpha(\tau)
    <
    \operatorname{LCB}_\alpha(\tau)
  \right\}.
\]
Collecting these events gives
$E=\bigcup_{\tau\in\mathcal{T}}E_\tau$, the event that at least one
candidate has an invalid lower confidence bound. The pointwise result in
Lemma~\ref{lem:app-pointwise-lcb}, together with the union bound, now gives
the following inequality. This bound does not require independence among
the events:
\begin{align}
  \limsup_{|D|\to\infty}\mathbb{P}(E)
  \leq
  \sum_{\tau\in\mathcal{T}}
  \limsup_{|D|\to\infty}\mathbb{P}(E_\tau)
  \leq
  \sum_{\tau\in\mathcal{T}}\frac{\delta}{|\mathcal{T}|}
  =
  \delta.
\end{align}
It follows that the complementary simultaneous-validity event satisfies:
\begin{align}
  \liminf_{|D|\to\infty}\mathbb{P}\!\left[
    \mathcal{B}_\alpha(\tau)
    \geq
    \operatorname{LCB}_\alpha(\tau),
    \ \forall\tau\in\mathcal{T}
  \right]
  \geq
  1-\delta.
\end{align}

Importantly, simultaneous validity does not require every candidate to have
a nonnegative LCB. The final selector considers only candidates satisfying
$\operatorname{LCB}_\alpha(\tau)\geq0$ and chooses the one with the largest
reduction. On the simultaneous-validity event, the selected candidate thus
satisfies
\(
  \mathcal{B}_\alpha(\tau^*)
  \geq
  \operatorname{LCB}_\alpha(\tau^*)
  \geq0.
\)
The F1-budget equivalence in Appendix~\ref{app:f1-budget} then implies
$\mathrm{F1}\geq\alpha$.
\end{proof}

\end{document}